\documentclass[fleqn,usenatbib]{mnras}

\usepackage{newtxtext,newtxmath}

\usepackage[T1]{fontenc}

\DeclareRobustCommand{\VAN}[3]{#2}
\let\VANthebibliography\thebibliography
\def\thebibliography{\DeclareRobustCommand{\VAN}[3]{##3}\VANthebibliography}

\usepackage{graphicx}	
\usepackage{amsmath}	

\title[Glasses for 2PCF estimates on light cones]{Glass-Like Random Catalogues for Two-Point Estimates on the Light Cone}

\author[]{Sebastian Schulz \thanks{E-mail: sebastian.schulz@uzh.ch}
\\
{Universit\"at Z\"urich, Institute for Computational Science, Winterthurerstr.\ 190, 8057 Z\"urich, Switzerland}}

\date{Accepted XXX. Received YYY; in original form ZZZ}

\pubyear{2023}

\begin{document}
\label{firstpage}
\pagerange{\pageref{firstpage}--\pageref{lastpage}}
\maketitle

\begin{abstract}
We introduce {\sc grlic}, a publicly available Python tool for generating glass-like point distributions with a radial density profile $n(r)$ as it is observed in large-scale surveys of galaxy distributions on the past light cone. Utilising these glass-like catalogues, we assess the bias and variance of the Landy-Szalay (LS) estimator of the first three two-point correlation function (2PCF) multipoles in halo and particle catalogues created with the cosmological N-body code \textit{gevolution}. Our results demonstrate that the LS estimator calculated with the glass-like catalogues is biased by less than $10^{-4}$ with respect to the estimate derived from Poisson-sampled random catalogues, for all multipoles considered and on all but the smallest scales. Additionally, the estimates derived from glass-like catalogues exhibit significantly smaller standard deviation $\sigma$ than estimates based on commonly used Poisson-sampled random catalogues of comparable size. The standard deviation of the estimate depends on a power of the number of objects $N_R$ in the random catalogue; we find a power law $\sigma \propto N_R^{-0.9}$ for glass-like catalogues as opposed to $\sigma \propto N_R^{-0.48}$ using Poisson-sampled random catalogues. Given a required precision, this allows for a much reduced number of objects in the glass-like catalogues used for the LS estimate of the 2PCF multipoles, significantly reducing the computational costs of each estimate.


\end{abstract}




\section{Introduction}

The large-scale spatial distribution of galaxies contains crucial information about the fundamental physics and the evolution of the Universe. For this reason, in the past decades substantial effort has been put into analysing its statistical properties in order to extract as much of that information as possible with the tools at hand \citep{Davis:1982gc, Vogeley:1992jx,Feldman:1993ky,Maddox:1996vz, 2dFGRS:2001ybp, Peacock:2001gs, SDSS:2003tbn, 2dFGRS:2005yhx, SDSS:2006lmn, Wang:2013noa, Shi:2016usd}. Common summary statistics of interest in these analyses are the two-point correlation function (2PCF) and its Fourier space counterpart, the power spectrum.

The 2PCF, commonly denoted as $\xi({\vec{d}})$, quantifies the excess probability of finding an object (e.g., a galaxy or a dark matter halo) at a position $\vec{x}+\vec{d}$ separated by $\vec{d}$ from another object at position $\vec{x}$, relative to the ensemble-average background number density $\bar{n}$:
\begin{equation}
     \xi(\vec{d}) = \langle \delta(\vec{x})\delta(\vec{x}+\vec{d})\rangle\,,
\end{equation}
where $\delta(\vec{x})$ is the number-density contrast of the objects,
\begin{equation}
\label{eq:delta}
    \delta(\vec{x}) = \frac{n(\vec{x})-\bar{n}}{\bar{n}}\,,
\end{equation}
with $n(\vec{x})$ the number density at position $\vec{x}$. In other words, the two-point correlation function of the galaxy distribution  captures its clustering properties by quantifying the correlation between pairs of galaxies at different separations.

Specifically, the 2PCF has been used to extract the comoving length scale of the sound horizon during the cosmic period known as matter-radiation decoupling. At that time, baryon-acoustic oscillations (BAO) were imprinted into the matter distribution (and consequently into today's galaxy distribution). This imprint is measurable in the galaxy 2PCF in the form of a peak at $\approx 100\,\mathrm{Mpc}/h$, which can act as a ``standard ruler'' and is hence useful for understanding the expansion history of the Universe \citep{SDSS:2005xqv, Blake:2012pj, SDSS:2009ocz, BOSS:2016wmc, eBOSS:2020yzd}.

There is more information to be extracted from the 2PCF by taking into account redshift space distortions (RSD) that mainly arise from the peculiar motion of the galaxies in their gravitational potential well \citep{Kaiser:1987qv}. The RSD then provide a way for measuring the growth rate of structure \citep{Guzzo:2008ac,Beutler:2012aaaaa,Reid:2012sw,delaTorre:2013rpa,Pezzotta:2016gbo,Zarrouk:2018vwy,Hou:2018yny,Ruggeri:2018kdn}, which in turn provides insights into the laws of gravity on large scales.

Future galaxy surveys such as the space based Euclid survey \citep{Euclid:2011zbd, Amendola:2016saw} or the ground based DESI survey \citep{DESI:2016fyo, DESI:2019jxc} aim to increase the accuracy and precision of the inferred cosmological parameters by quantifying the clustering statistics of galaxies in much larger volumes with accurate measurements of the redshifts of tens of millions of galaxies.

In order to estimate the galaxy 2PCF, it is generally assumed that the observed galaxy distribution represents a Poisson sample of some underlying galaxy density field \citep{1980lssu.book.....P}. Given the point distribution of observed galaxies, it is natural to use estimators based on pair counts. Many estimators of this kind have been proposed over the years \citep{Peebles:1974aaaaa, Hewett:1982aaaaa, Davis:1982gc, Hamilton:1993fp, Landy:1993yu}. Among those, the most commonly used for the 2PCF of large-scale structure is the Landy-Szalay (LS) estimator \citep{Landy:1993yu}, which is defined in its general form for the cross-correlation between two data catalogues $D_1$ and $D_2$
\begin{equation}
\label{eq:ls}
\xi_\mathrm{LS}(d,\mu) = \frac{\widehat{D_1 D_2}(d,\mu) - \widehat{D_1 R_2}(d,\mu) - \widehat{R_1 D_2}(d,\mu) + \widehat{R_1 R_2}(d,\mu)}{\widehat{R_1 R_2}(d,\mu)}\,,
\end{equation}
introducing two random catalogues $R_1$ and $R_2$ with the same radial density profile and survey mask as the corresponding data catalogues. For autocorrelations within one data catalogue, $D_1$ and $D_2$ are identical. In Eq.~\eqref{eq:ls}, $d$ is the absolute value of the separation between a pair of objects within the catalogues and $\mu$ is the cosine of the angle between the line of sight to the pair $\vec{s}$ and the separation vector $\vec{d}$. The $\widehat{D_1D_2}$, $\widehat{D_1R_2}$, $\widehat{R_1D_2}$ and $\widehat{R_1R_2}$ correspond to histograms of pair counts between the data ($D_i$) and random ($R_i$) catalogues, binned in $d$ and $\mu$ and normalised by the total number of pairs between the respective two catalogues. In general, the normalised histogram of pair counts between two catalogues $A$ and $B$ is given by


\begin{equation}
\label{ls_norm}
    \widehat{AB}(d,\mu) = \frac{AB(d,\mu)}{N_{A}N_{B}-\frac{1}{2}\delta_{AB}N_{A}(1+N_{A})}\,,
\end{equation}
where in our case $A,B\in\{D_1,D_2,R_1,R_2\}$. Here, $AB(d,\mu)$ is the number of pair counts between catalogue $A$ and catalogue $B$ in the respective $(d,\mu)$ bin, $\delta_{AB}$ is a Kronecker-Delta which is 1 if $A=B$ and 0 if $A\neq B$. Hence, for cross-correlations of two different data catalogues $D_1$ and $D_2$, the $\delta_{AB}$ term vanishes, but for autocorrelations, where $A=B$, it is included because in this case pairs are usually not double-counted. The same arguments apply for the $R_iR_j$ pairs, while for the $D_iR_j$ pairs the $\delta_{ij}$ term always vanishes, since those pairs are always counted between two separate catalogues, $A\neq B$.

When the correlations are small ($\xi \ll 1$), which is true for the galaxy distribution of the Universe on large scales, the LS estimator has the lowest bias and variance of all possible estimators based on pair counts if a very densely sampled random catalogue which follows the same redshift distribution as the data catalogue is provided \citep{Landy:1993yu,Kerscher:1999hc,Keihanen:2019vst}. In this case, the variance of the estimator is almost Poisson.
Commonly the number of objects in the random catalogue exceeds the number of objects in the data catalogue by a factor of order ten or higher to ensure this criterion is fulfilled in order to keep the bias and variance small \citep{Samushia:2011cs, delaTorre:2013rpa, BOSS:2016off, Bautista:2020ahg}.
We note that at smaller scales (below $\approx 10\,\mathrm{Mpc}/h$), the galaxy 2PCF can become of order one, in which case the LS estimator is not the optimal estimator anymore. To account for this, \cite{Vargas-Magana:2012szx} have derived a new optimal estimator that remains unbiased on smaller scales as well.

The random catalogue serves to approximate the pair counts in a homogeneous distribution, which are then used together with the pair counts in the data distribution and the pair counts across data- and random catalogues to derive an estimator for the 2PCF. A common choice for the random catalogue for large-scale galaxy surveys is to Poisson-sample from a distribution that is uniform in angular space and follows the redshift distribution of the data. A survey mask is taken into account by setting the number of objects in the random catalogue to zero outside of the mask.

We note that commonly the uncertainty in the 2PCF for galaxy clustering observations is dominated by sample variance, originating from the fact that the 2PCF in one region of the sky can differ from the 2PCF in another region of the sky, and from shot noise due to the data points being Poisson-sampled from the underlying density distribution. Here we focus instead on the uncertainty in the estimator arising from the statistical fluctuations in the number of data-random and random-random pairs, neglecting the sample variance. Keeping the variance of the estimator as small as possible is desired, so that the only remaining dominant contribution to the variance is the sample variance. 

The variance of the LS estimator is dominated by the density fluctuations of the random catalogue \citep{Keihanen:2019vst}. The amplitude of these fluctuations is given by the power spectrum, which, in case of a Poisson-sampled catalogue, is $P(k)=1/\bar{n}$, where $k = 2\pi/\lambda$ is the absolute value of the wave vector of the Fourier-space fluctuations with wavelength $\lambda$. The variance within a sphere of radius $R$ is related to the power spectrum \citep{Gabrielli:2001xw}:
\begin{equation}
\sigma^2(R) = \frac{1}{{2\pi}^2} \int P(k) W^2(kR) k^2 dk\,,
\end{equation}
where $W(kR)$ is the Fourier transform of a top-hat window function,
\begin{equation}
W(kR) = 3 \frac{\sin(kR)-kR\cos(kR)}{(kR)^3}\,.
\end{equation}

For the Poisson sample, which has a constant power spectrum, the variance goes like $\sigma^2(R)\propto R^{-3}$. Increasing the number of objects in the random catalogues will decrease the amplitude of the power spectrum and thereby the variance of the LS estimator, but this comes at a significant additional computational cost as the pair-counting process goes like $\mathcal{O}(N^2)$. With ever increasing catalogue sizes approaching hundreds of millions of galaxies, having random catalogues with many more objects than the data catalogues will eventually become unfeasible. Since most of the computation time for the estimate is taken up by the counting of the random-random pairs, it is promising to look for approaches that reduce the time taken to perform this computation given a desired precision of the estimator, for example by requiring smaller random catalogues, or none at all.

\cite{Keihanen:2019vst} propose a method to increase the speed of the $R_iR_j$ calculation by splitting the random catalogue $R$ into $M_s$ sub-catalogues $R^\mu$ and then taking the average of the normalised pair-counts within each sub-catalogue. This method does not affect the accuracy of the estimator and can therefore be used to significantly speed up the computation time. However, the cost can still become high if the survey is very large or if the estimator has to be calculated for a large number of data catalogues.

\cite{Breton:2020ydo} provide a scheme to analytically calculate the pair counts involving the random catalogue, given the survey window function and the radial selection function, speeding up the estimate significantly. The method requires an estimate of the 2PCF of the angular selection function beforehand. An analytical method for higher-order statistics is yet to be derived, making random catalogues necessary for estimates of $(N>2)$-point correlation functions.

Alternatively, it is worthwile exploring random catalogue configurations that have a lower variance than the Poisson-sampled catalogues. The fastest decay of the variance with increasing $R$ for any statistical point distribution in three dimensions is $\sigma^2(R) \propto R^{-4}$ which is achieved by a class of distributions that exhibit smaller power on large scales than the uniform Poisson distribution with a power spectrum $P(k) \propto k^4$. These distributions are called super-homogeneous and examples include periodic grids or glasses \citep{Gabrielli:2001xw}. 

Glasses have smaller power than a Poisson distribution on scales larger than the average inter-particle separation, which is why they have been of interest in the scientific literature in the past, but in a slightly different context. It was suggested that these kind of distributions can be used to generate pre-initial conditions for cosmological simulations, which need to have as little power as possible \citep{White:1994bn, Baugh:1994hb, Hansen:2006hj, Joyce:2008kg}. However, it was later found that glasses are not superior to periodic crystals in this regard. Therefore glasses are less commonly used in today's cosmological simulations because of the difficulty that lies in creating them.

Recently, \cite{Davila-Kurban:2020eph} have shown that using glass-like catalogues leads to a smaller variance in the LS estimates of the 2PCF in comparison to using Poisson random catalogues. They demonstrated this on a mock catalogue of simulated galaxies on an equal-time hypersurface in a periodic box. They found that an LS estimate with a desired precision would require glass catalogues with fewer objects than what would be needed using Poisson-sampled random catalogues.

In general, glass distributions can be created from a uniform Poisson distribution by running an N-body simulation with repellent gravitational forces, but this requires a lot of computing resources. As an alternative, \citet{Davila-Kurban:2020eph} have suggested a fast way of generating glass-like catalogues using the Zeldovich approximation of Lagrangian perturbation theory \citep{Zeldovich:1970aaaaa}. Starting with a Poisson sample of a uniform distribution in a periodic simulation box, one can iteratively displace the objects in a direction opposite to the one given by the first-order displacement field $\vec{\Psi}$. After many iterations, the power spectrum of the catalogue is approximately the one of a glass, with $P(k)\propto k^4$. In their work, \citet{Davila-Kurban:2020eph} adapted the publicly available BAO reconstruction code of \citet{Bautista:2017wwp}\footnote{\url{github.com/julianbautista/eboss_clustering}} based on the Fourier-space algorithm of \citet{Burden:2015pfa}.

In this article, we aim to extend this approach to catalogues on the light cone with a redshift-dependent comoving number density of objects $n(r(z))$, which can be due to a radial survey selection function or due to an intrinsic redshift-dependent abundance of the objects (for example, high mass dark matter halos are less abundant at high redshifts). We provide the publicly available Python tool {\sc grlic}, to easily create glass-like random catalogues given the survey specifications (opening angle, survey mask, redshift range and the redshift dependent number density $n(r(z))$). We stress the fact that the glass-like catalogues might not be considered truly ``random'' due to the correlated nature of their data points. Nevertheless, each glass-like catalogue is a random realisation of the underlying background distribution and we will occasionally continue to use this term to underscore the glass-like catalogues' function as the reference ``random'' dataset in the pair-count estimations of the 2PCF.

Glasses are particularly useful in this context as they allow for a smooth evolution of the number density with radial distance, without any discontinuities as they would be unavoidably present in a crystal-like structure. 

We use the glass-like random catalogues in the LS estimator to estimate the multipoles of the two-point auto- and cross-correlation functions of a set of simulated catalogues of dark matter particles and halos and find that the resulting estimates using the glass-like random catalogues have a much smaller variance than the estimates using a Poisson-sampled random distribution. There is effectively no bias with respect to the result obtained using a much larger Poisson-sampled random catalogue with an identical $n(r(z))$.

The paper is structured as follows: in Sec.~\ref{sec:Methods} we describe the methods used to extract the radial density profile from a data catalogue, to Poisson sample a random catalogue with an identical radial density profile and to generate the glass catalogues from them. The application of the Zeldovich method is explained in detail and the measurement procedure for the 2PCF multipole estimates is described. In Sec.~\ref{sec:Data} we describe the simulations used to create the data catalogues, the data catalogues themselves, and the the glass catalogues used in the LS estimates. In Sec.~\ref{sec:results} we present our results on the estimates of the 2PCF multipoles, as well as measurements of their bias and standard deviation. Sec.~\ref{sec:conclusion} contains a summary of this work and a final discussion of the results. Appendix~\ref{app:convergence} contains convergence tests of our fiducial setup for creating the glass-like random catalogues. Appendix~\ref{app:mask} contains results on an estimate of the 2PCF after applying an exemplary mask to the catalogues.

\section{Methods for creating glass catalogues on the light cone}
\label{sec:Methods}

The goal of the following pipeline is to create a glass-like catalogue that mimics the redshift-dependent comoving number density $n(r(z))$ of a given data catalogue. The resulting glass has to be locally isotropic in comoving coordinates, which is why it is convenient to first convert the observed redshifts $z$ of the data catalogue into comoving distances $r$ assuming a fiducial background cosmology with specified cosmological parameters. Then, the comoving number density is derived from a histogram of the number counts $N(r)$, taking the survey geometry into account. 

Alternatively, the user of the code can directly provide a tabulated $n(r(z))$ together with the cosmological parameters that fix the mapping between $r$ and $z$.

We then Poisson sample a random catalogue that follows the $n(r(z))$ of the data catalogue, and iteratively displace the objects in comoving coordinates until they reach a glass-like distribution. The steps will be outlined in detail in the following subsections. 

\subsection{Extracting $n(r)$ from the data catalogue}
Taking into account that galaxy surveys probe times well into matter domination, the observed redshifts in the data catalogue are converted into comoving distances according to

 \begin{equation}
     r(z) \approx \int_0^z \frac{c \mathrm{d}z'}{H_0 \sqrt{\Omega_m (z'+1)^3+\Omega_\Lambda + (1-\Omega_m-\Omega_\Lambda) (z'+1)^2}}\,,
 \end{equation}
where $c$ is the speed of light, $H_0 \equiv 100\,h\,\mathrm{km}\,\mathrm{s}^{-1}\mathrm{Mpc}^{-1}$ the value of the Hubble constant today, $\Omega_m$ the density parameter for the matter content in the Universe today and $\Omega_\Lambda$ the density parameter of dark energy today. In the remainder of this work we will implicitly assume that the inferred comoving distance $r(z)$ depends on the observed redshift $z$ and drop the argument of the function, $r(z) \to r$.
Note that redshift-space distortions (RSD) are not taken into account here, as the goal is to derive the large-scale $n(r)$ behaviour, which will not be affected by RSD significantly due to continuity within the survey. 

The number of objects $N(r)$ within a bin of width $\Delta r$ centered at the comoving distance $r$ is related to the observed number density $n(r)$ via

\begin{equation}
\label{eq:N(r)}
    N(r) = \int^{r+\Delta r/2}_{r-\Delta r/2} n(r') r'^2 \int_0^\pi \sin{\theta} \int_0^{2\pi} W(r',\theta,\phi)\,\mathrm{d}r'\mathrm{d}\theta\,\mathrm{d}\phi\,,
\end{equation}
where $W(r,\theta,\phi)$ is the survey window function that generally depends on $r$, the polar angle $\theta$ and the azimuthal angle $\phi$. In the remainder of this work, we will assume simple survey shapes for light cones with $W(r_\mathrm{min} \leq r \leq r_\mathrm{max}, \theta \leq \theta_\mathrm{max},\phi) = 1$ and $W=0$ elsewhere. In this case, $\theta_\mathrm{max}$ is the half-opening angle of the survey and $r_\mathrm{min}$ and $r_\mathrm{max}$ are the nearest and farthest inferred distances of any object in the survey to the observer. The window function can easily be generalised to more complex survey shapes. The number density within each bin can then be estimated from the number of objects in it:
\begin{equation}
\label{eq:n(r)}
n(r) \approx \frac{N(r)}{2\pi r^2 (1-\mu_\mathrm{min}) \Delta r}\,,
\end{equation}
where we assumed that $\Delta r$ is small compared to $r$ and $\mu_\mathrm{min} \equiv \cos{\theta_\mathrm{max}}$.

\subsection{Poisson sampling a random catalogue}
\label{sec:poisson_randon}
The purpose of the random catalogue used within pair-count estimators such as the LS estimator is to serve as a sample of a homogeneous distribution that mimics the overall radial distribution of the data catalogue. In general the $n(r)$ measured from the data will exhibit fluctuations around the underlying background evolution (which is the quantity we are interested in) due to galaxy clustering, and these fluctuations become more pronounced as the half-opening angle of the survey gets smaller (effectively this is due to increasing cosmic variance). Hence, in order to keep the information to be gained from the clustering along the line of sight in the estimate, smoothing the $n(r)$ is desirable. The analytical form of this function is not generally defined, as it depends on the survey selection function and on the intrinsic redshift evolution of the abundance of the observed objects, which in turn depends on the physical properties of the objects, i.e. their masses, luminosities, or colours. Within this work, we perform a third-order polynomial fit to the measured $n(r)$, to model the overall background evolution of the number density of ``observed'' dark-matter halos (see Sec.~\ref{sec:Data} for more details).

We then Poisson sample a random catalogue from the $n(r)$ distribution using the inversion method, where random numbers are drawn from a uniform distribution between zero and one and converted into a distance $r$ by inverting the cumulative probability distribution, 

\begin{equation}
    P_\mathrm{cumul}(r) = \frac{2\pi (1-\mu_\mathrm{min})}{N} \int_{r_\mathrm{min}}^r n(r') r'^2 dr'\,,
\end{equation}
where $N$ is the total number of objects in the data catalogue. Sampling $N_R = \alpha N$ objects results in a density profile $\alpha n(r)$.
In the case of the third-order polynomial fit to $n(r)$, the integral for the cumulative probability can be easily calculated analytically. If the functional form of the $n(r)$ is not known, e.g. if it is simply a result of smoothing the measured $n(r)$ with some kernel, the integral can be performed numerically. We interpolate such $n(r)$ with a cubic spline kernel to achieve high numerical precision in the integral.

In general the functional form of $P_\mathrm{cumul}$ is not known, and even if it is known, finding the inverse is not always trivial. Hence we simply find the mapping between $r$ and $P_\mathrm{cumul}$ using again a cubic spline interpolation.

We randomly sample the  $\cos{\theta}$ from a uniform distribution between $\mu_\mathrm{min}$ and $1$, and the $\phi$ from a uniform distribution between $0$ and $2\pi$. These Poisson-sampled randoms are then the initial catalogues used for creating the glass-like random catalogues.

\subsection{From Poisson random catalogues to glass-like random catalogues}

Starting from an initial set of Poisson-sampled objects distributed randomly and uniformly in a periodic box one can create a glass by iteratively displacing the objects with a repellent force acting between each pair of objects until an equilibrium distribution is reached. It is convenient to assign some mass to the objects and simply use the force of gravity acting in the opposite direction. A full N-body simulation would be the most accurate implementation of this displacement, but this requires additional computing resources which might become too large in comparison to the eventual gain in accuracy of the estimator. Hence, in this work we employ the Zeldovich approximation as suggested by \citet{Davila-Kurban:2020eph}.

While this is more or less straightforward to implement for distributions with a uniform number density, i.e. $n(r) = \mathrm{const.}$, for a non-uniform $n(r)$ additional steps are required to include the effects of external forces that force the objects to take the desired distribution. Those steps will be outlined below.

\subsubsection{Zeldovich approximation}
\label{sec:zel}
Lagrangian perturbation theory models the gravitational dynamics on large scales in the Universe, where matter density perturbations are small, to high accuracy. A central quantity in this theory is the displacement field $\vec{\Psi}$, which relates the initial Lagrangian position of a fluid element $\vec{q}$ to Eulerian positions $\vec{x}$ observed at time t,

\begin{equation}
\vec{x}(\vec{q},t) = \vec{q} + \vec{\Psi}(\vec{x},t)\,.
\end{equation}
Assuming mass conservation between the Eulerian and Lagrangian frame, the displacement field can then be related to the Eulerian density perturbations. The linear-order solution of the displacement field is
\begin{equation}
\vec{\nabla}_q \cdot \vec{\Psi}_1(\vec{q},t) =-\delta_1(\vec{x},t)\,,
\end{equation}
where $\delta_1$ is the linear contribution to the density contrast defined in Eq.~\eqref{eq:delta} and $\vec{\nabla}_q$ is the differential nabla operator in Lagrangian coordinates, i.e. $\vec{\nabla}_q \cdot \vec{a}$ gives the divergence of the vector field $\vec{a}$. 

The displacement field is solved for in Fourier space:
\begin{equation}
\label{eq:displacement}
\vec{\Psi}_1(\vec{k},t) = \frac{i\vec{k}}{k^2}\delta_1(\vec{k},t)\,,
\end{equation}
where we use the following Fourier convention:
\begin{equation}
\tilde{f}(\vec{k})=\int d^3x f(\vec{x}) e^{-i\vec{k}\cdot\vec{x}}\,,
\end{equation}
\begin{equation}
f(\vec{x})=\frac{1}{(2\pi)^3}\int d^3k \tilde{f}(\vec{k}) e^{i\vec{k}\cdot\vec{x}}\,.
\end{equation}

Switching the sign of the displacement field effectively mimics a situation with repellent gravitational forces. This can be used to create a glass-like distribution from an initial Poisson distribution, if such a displacement is applied iteratively until an equilibrium configuration is reached, where all the repulsive forces are balanced. 

Numerically this procedure is implemented as follows. First, the domain is discretised into a sufficiently fine grid and the masses of the particles are assigned to the grid cells with a mass assignment scheme, e.g. one of the ``nearest grid point'', ``cloud in cell'' or ``triangular shaped cloud'' schemes. The density contrast within each grid cell is estimated according to Eq.~\eqref{eq:delta} and then Fourier transformed using a fast Fourier transform (FFT) algorithm. The Fourier-space displacement field in each grid cell is then estimated from the linear density contrast in each grid cell following Eq.~\eqref{eq:displacement}, and transformed back with an inverse FFT. Finally the displacement field is interpolated at the particle positions with a scheme consistent with the chosen mass-assignment scheme. Its negative value is then used to update the particle positions. The density contrast within each cell is updated and the process is repeated until the displacements are sufficiently small.

If a given redshift-dependent $n(r)$ distribution is to be reproduced, it can be incorporated into the background number density from which the density contrast is calculated, $\bar{n} \rightarrow \bar{n}(r)$, i.e. the density contrast in each grid cell is calculated with respect to the desired radially dependent number density of the glass-like random catalogue.

In order to achieve periodic boundary conditions, the observer is placed in the center of the box, i.e. the distance $r$ to the center of the box is set to zero, and the box size is chosen so that its side length is at least twice the maximum comoving distance $r_\mathrm{max}$ in the survey. A buffer zone is added for distances within the box that are larger than $r_\mathrm{max}$, with constant background number density $\bar{n}(r>r_\mathrm{max}) = n(r_\mathrm{max})$  which ensures smooth and differentiable boundary conditions required for the FFT within the periodic box. Grid cells that are located at $r < r_\mathrm{min}$ are assigned a background number density of $\bar{n}(r<r_\mathrm{min}) = n(r_\mathrm{min})$. 

We note that the so-defined $\bar{n}(r)$ is not differentiable at $r_\mathrm{min}$ and $r_\mathrm{max}$, but as long as $n(r)$ evolves reasonably weakly within the survey, this does not cause strong edge effects when later Fourier transforming the number densities on the grid, as the non-differentiability gets smoothed out by the sampling of the background number density on the grid. For stronger evolutions, the $\bar{n}(r)$ can be extrapolated to $r$ that lie sufficiently outside of the survey, so that the resulting glass catalogue does not suffer from edge effects near $r_\mathrm{min}$ and $r_\mathrm{max}$. However, this requires increasing the size of the buffer zone to ensure smoothness on the boundaries, which in turn increases the required grid resolution to achieve the same resolution within the actual survey volume. 

The glass is then initialised by Poisson sampling a random catalogue as outlined in Sec.~\ref{sec:poisson_randon} within the full spherical survey shell, i.e. setting $\mu_\mathrm{min} = -1$, and Poisson sampling uniformly from $n(r < r_\mathrm{min}) \equiv {n}(r_\mathrm{min})$ and, respectively, $n(r>r_\mathrm{max}) \equiv {n}(r_\mathrm{max})$ in the buffer zones. We then iteratively displace the objects with the reversed Zeldovich displacement field $N_\mathrm{iter}$ times with the goal to achieve a glass-like structure. After $N_\mathrm{iter}$ iterations, only the objects within the specified survey volume are kept in the final glass-like catalogue.


Depending on the Fourier grid resolution, after a certain amount of iterations, the objects will start to align with the grid, effectively introducing long-range periodicities in the distribution,  which is to be avoided if the goal is to create a non-periodic glass. Hence, a sweet spot is to be found, where each iteration reduces the initially Poisson variance, while too many iterations lead to a periodic crystal-like structure. Again, such a crystal-like structure is undesirable for the task at hand, because it introduces artificial periodicities and anisotropies into the random catalogue.




\subsection{Measuring the 2PCF}
The LS estimate of the 2PCF is perfomed according to Eq.~\eqref{eq:ls}, where we use a mid-point line of sight definition: $\vec{s}  = \vec{x}+\vec{d}/2$. Eq.~\eqref{eq:ls} can be used for the cross-correlation between two sets of objects, $D_1$ and $D_2$, as well as for the autocorrelation by substituting $D_2$ and $R_2$ with $D_1$ and $R_1$.

We note that in the case of autocorrelations, the glass approach requires using two separate glass catalogues, since the data points within the glass are now correlated. Using only one glass catalogue would lead to a biased estimate of the $R_1R_2(d,\mu)$ \citep{Davila-Kurban:2020eph}. This is not a necessary step when estimating autocorrelations using a Poisson-sampled random catalogue, but since using two separate random catalogues lowers the variance of the estimator \citep{Davila-Kurban:2020eph}, we will also use two separate Poisson-sampled random catalogues for fair comparisons of variances of the autocorrelation estimates using glasses and Poisson samples.

In this work we are using a modified version of the publicly available code {\tt{CUTE}} \citep{Alonso:2012rk}, that implements the described LS-estimator in a very efficient way making optimal use of parallel computing. The modifications follow the procedure of \cite{Breton:2018wzk}, which allows for the measurement of odd multipoles in the full 3D correlation function. We note that we use the cross-correlation algorithm of {\tt{CUTE}} even for the autocorrelations, meaning that we double count the pairs in the data catalogues and hence set $\delta_{ij}$ to zero in Eq.~\eqref{ls_norm}.

The multipoles are then estimated from the full correlation function according to

\begin{equation}
\label{eq:multipole}
    \xi_\ell(d) \approx \frac{2\ell+1}{2} \sum^1_{\mu=-1} \xi_\mathrm{LS}(d,\mu) \mathcal{L}_\ell(\mu) \Delta\mu\,,
\end{equation}
where $\mathcal{L}_\ell(\mu)$ is the $\ell$-th order Legendre polynomial and $\Delta\mu$ is the width of the $\mu$-bin. 

\subsection{Variance of the LS estimator}
\label{sec:variance}
Let us reconsider the LS estimator of Eq.~\eqref{eq:ls}. The bias and variance of the LS estimator has been derived in detail in \cite{Landy:1993yu}, assuming an infinitely large random catalogue. In this limit, and with infinitesimally small bins, the estimator is unbiased and has almost Poisson variance. In \cite{Keihanen:2019vst} the derivation was generalized to finite random catalogues, which introduces a small bias to the estimate. While the variance of the LS estimate depends on the size of the bins used for the pair counts, as well as on geometric properties of the survey (these are the so-called edge terms in the variance), to zeroth order in $\xi$ the LS variance is dominated by the Poisson variances of the pair counts. Then, keeping the number of objects in the data catalogues fixed, the biggest contribution to the variance of the LS estimate comes from the $D_iR_j$ terms, as their Poisson variance decreases as $\sigma^2(D_iR_j) \propto N_{R_j}^{-1}$. The contribution from the $R_1R_2$ term is suppressed, as it has a variance that decreases more quickly, $\sigma^2(R_1R_2) \propto N_{R_1}^{-1}N_{R_2}^{-1}$. For very large numbers of objects in the Poisson-sampled random catalogue, the variance of the LS estimator is hence expected to be almost proportional to $N_{R_j}^{-1}$. 

Using glass-like catalogues instead of Poisson-sampled catalogues leads to a different expected variance of the LS estimator. We will assume that its variance is still dominated by the variance of the $D_iR_j$ terms, which is now no longer Poisson. As discussed before, the variance within a sphere of radius $R$ scales like $R^{-4}$ for a glass distribution, in contrast to the $R^{-3}$ scaling of a Poisson distribution. Increasing $R$ at fixed number density of objects is equivalent to increasing the number density of objects within a sphere at fixed $R$. From this argument and from $N_{R_j} \propto R^3$ it follows that in the limit of very large $N_{R_j}$ the variance of $D_iR_j$ approximately evolves like $N_{R_j}^{-1}$ if the $R_j$ catalogue is Poisson-sampled, and it approximately evolves like $N_{R_j}^{-4/3}$ for a glass catalogue. If the number of objects in the random catalogue is much larger than the number of objects in the corresponding data catalogue by a factor of $\alpha$, i.e. $N_{R_j} = \alpha N_{D_j}$, then we expect the standard deviation $\sigma$ of the LS estimate to decrease approximately like $\sigma \propto \alpha^{-0.5}$ for the estimate utilising the Poisson-sampled random catalogues, and like $\sigma \propto \alpha^{-2/3}$ for the estimate derived from the glass-like catalogues.
\section{Data catalogues}
\label{sec:Data}
\begin{figure}
    \includegraphics[width=\columnwidth]{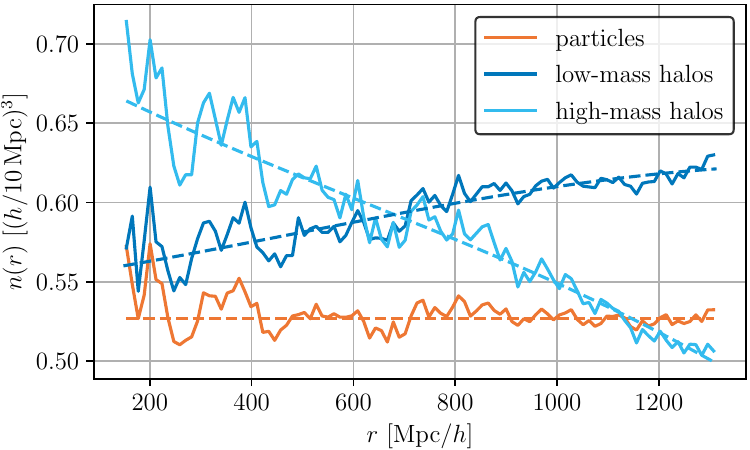}
    \caption{Number density of objects in the particle and halo catalogues studied in this work. The solid lines represent the number of objects per $(10\,\mathrm{Mpc}/h)^3$ within spherical shells of thickness $\Delta r \approx 12\,\mathrm{Mpc}/h$. The orange dashed line represents the constant average number density of the particles, while the blue dashed line represents the third-order polynomial fits to the $n(r)$ curves of the halo catalogues.}
    \label{fig:n_of_r}
\end{figure}

To test the validity of our approach for generating glass-like random catalogues with a specified $n(r)$ distribution, we estimate the two-point autocorrelation and cross-correlation functions for a set of particle and halo light cones and compare the bias and variance of the estimators using different numbers of objects within the glass and random catalogues. The particle light cone used in this work is generated with \textit{gevolution}, a relativistic N-body code for cosmological simulations which produces light-cone data in comoving space during run time \citep{Adamek:2016zes}. Notably, the full space-time metric on the light cone is stored, allowing for self-consistent ray tracing of the objects on the light cone. In this work, we make use of the full-sky particle light cone of the \texttt{unity2-lowz} simulation, with a number of $N_\mathrm{part} = 5760^3$ particles within a periodic box of volume $V_\mathrm{box} = (4032\,\mathrm{Mpc}/h)^3$. This translates to a mass resolution of $M_\mathrm{part} = 3 \times 10^{10}\,M_\odot/h$. Within \textit{gevolution} the force of gravity acting on each particle is calculated with a particle-mesh scheme -- the mesh resolution within the \texttt{unity2-lowz} simulation is $700\,\mathrm{kpc}/h$, i.e. the number of grid cells is exactly the number of particles in the periodic simulation box. The underlying cosmology is a standard $\Lambda$CDM model with three neutrino species with masses of $0\,\mathrm{eV}$, $0.008689\,\mathrm{eV}$ and $0.05\,\mathrm{eV}$, respectively. The cosmological parameters are set to $A_{s} = 2.215\times 10^{-9}$, $n_s = 0.9619$, $h = 0.67$, $\Omega_\mathrm{b} = 0.049$, $\Omega_\mathrm{cdm} = 0.26858$ and $T_\mathrm{CMB} = 2.7255\,K$. Adding the massive neutrino contribution to the total matter density, we end up with $\Omega_m = 0.31898$ and $\Omega_\Lambda = 0.68095$, i.e. the global spacetime is flat. The initial conditions are generated from the linear transfer functions of {\sc class} \citep{Blas:2011rf}.
\begin{table}
	\centering
	\caption{Summary of data catalogues used in this work. All catalogues are light cones on the full sky with a redshift range of $0.05 \leq z < 0.5$.}
	\label{tab:catalogues}
	\begin{tabular}{lll} 
		\hline
		object type & number of objects & number of particles per object\\
		\hline
		particle & $4\,999\,574$ & $N_\mathrm{part} = 1$\\
            low-mass halo & $5\,759\,987$ & $30 \leq N_\mathrm{part} < 40$\\
		high-mass halo & $5\,204\,348$ & $N_\mathrm{part} \geq 300$\\
		\hline
	\end{tabular}
\end{table}
In \citep{Lepori:2022hke}, dark matter halos are identified as spherical overdensities by running the {\sc rockstar} halo finder \citep{Behroozi:2011ju} on the particle light cone in comoving space. The halos and particles are then raytraced through the metric light cone using the ray-tracing algorithm described in \citet{Lepori:2020ifz} to finally obtain a full-sky particle catalogue and a full-sky halo catalogue on the light cone in redshift space, including RSD and other relativistic effects. In this work, we make use of those catalogues. Specifically we select a subset of particles and halos from the full light cone outputs at redshifts $0.05 \leq z < 0.5$, which translates to $r_\mathrm{min} \approx 148\,\mathrm{Mpc}/h$ and $r_\mathrm{max} \approx 1313\,\mathrm{Mpc}/h$. For the particle light cone we randomly select $\sim 5 \times 10^6$ objects and from the halo catalogue we extract two subsamples: one that contains low-mass halos that are each made up of $30 \leq N_\mathrm{part} < 40$ FoF particles as identified by {\sc rockstar}, corresponding to an FoF mass range of $9\times 10^{11} M_\odot/h \leq M_\mathrm{FoF} < 1.2\times 10^{12} M_\odot/h$, and one that contains high-mass halos that are made up of $N_\mathrm{part} \geq 300$ particles each, corresponding to masses $M_\mathrm{FoF} \geq 9 \times 10^{13} M_\odot/h$. The catalogues used in this work are summarised in Table \ref{tab:catalogues}. For the halo catalogues, we fit a polynomial of third order to the $n(r)$, weighting the fit by the inverse of the Poisson error within each shell, i.e. we minimise
\begin{equation}
    E=\sum^N_{i=0} |\frac{n_i-\hat{n}}{\sigma_i}|^2\,,
\end{equation}
where $n_i$ is the number density within the $i$-th $r$-bin, $\hat{n}$ the value of the polynomial at the $i$-th $r$-bin and $\sigma_i$ the standard deviation of the number density within the $i$-th $r$-bin, which is assumed to be approximately Poisson,
\begin{equation}
    \sigma_i = \frac{\sqrt{N_i}}{V_i}\,,
\end{equation}
where $N_i$ is the number of objects in the $i-th$ bin and $V_i$ is the volume of the $i$-th shell with width $\Delta r$.
For the particle light cone, we assume $n(r)$ to be constant, as the number of dark matter particles is conserved in the simulation and there is no selection function for our survey, i.e. all objects are observed with a probability of $P=1$. The number density of the particles $n(r) = \mathrm{const.} = \bar{n}$ is estimated from dividing the total number of particles in the catalogue by the survey volume. We plot the radially dependent number density of objects for each catalogue in Fig. \ref{fig:n_of_r}, together with the background $n(r)$ derived from the polynomial fits to the binned halo number densities and the constant value associated with the particle number density.

\subsection{Glass catalogue specifications}
\label{sec:glass_desc}
Following the steps outlined in Sec.~\ref{sec:Methods}, we create Poisson-sampled random catalogues and glass-like random catalogues with different numbers of objects $N_{R_i} = \alpha N_{D_i}$, where $\alpha\in\{0.5, 1,2,10, 20\}$. To be precise, for each data catalogue listed in Table~\ref{tab:catalogues} and each value of $\alpha$, we create 20 pairs of independent random catalogues $R_1$ and $R_2$ to be used for the LS estimate according to Eq.~\eqref{eq:ls}, giving 40 independent Poisson-sampled and 40 independent glass-like random catalogues per $\alpha$ per data catalogue. We note that the Poisson-sampled catalogues have exactly $N_{R_i} = N_{D_i}$ objects, while the glass-like catalogues can end up with more or fewer objects as some objects can be displaced across the survey volume boundaries during the Zeldovich iterations. The fluctuation in the number of objects in the glass-like catalogues is below one percent of the desired number.

We choose a number of Zeldovich iterations $N_\mathrm{iter}=2$, the impact of this choice is tested by creating additional glass-like catalogues for the high-mass halo catalogue, using numbers of iterations $N_\mathrm{iter}\in\{1,2,3,5,10\}$ in Appendix~\ref{app:convergence}. The box size is chosen such that the buffer zone spans $\sim 400\,\mathrm{Mpc}/h$ on both sides of each direction, i.e. the side length of the cubic box in which we evolve the glasses is $L_\mathrm{box} = 2\times(r_\mathrm{max} + 400)\,\mathrm{Mpc}/h \approx 3427\,\mathrm{Mpc}/h$. Following the approach of \citet{Davila-Kurban:2020eph} the number of grid points is chosen such that the cell size is approximately one quarter of the average inter-particle separation in the light cone, i.e. for the particle light cone, with $n \approx 0.5\,(h/\mathrm{10\,Mpc})^3$ the average inter-particle separation is $d_\mathrm{inter}=n^{-1/3}\approx12.6\,\mathrm{Mpc}/h$. 
This leads to a required number of $N_\mathrm{grid} \approx 1080$ grid cells in one grid dimension, i.e. the total number of grid cells is $(N_\mathrm{grid})^3$. In order to speed up the FFT, we choose the nearest number that is a power of two, which results in $N_\mathrm{grid} = 1024$. We test the impact of this choice by varying the number of grid cells $N_\mathrm{grid} \in \{128, 256, 512, 1024\}$ in Appendix~\ref{app:convergence}. We employ a ``cloud in cell'' mass assignment scheme that is implemented in the public code {\sc Pylians} \citep{Pylians} to distribute the particle point masses onto the grid and to interpolate the displacement field at the particle positions.

\begin{figure}
    \includegraphics[width=\columnwidth]{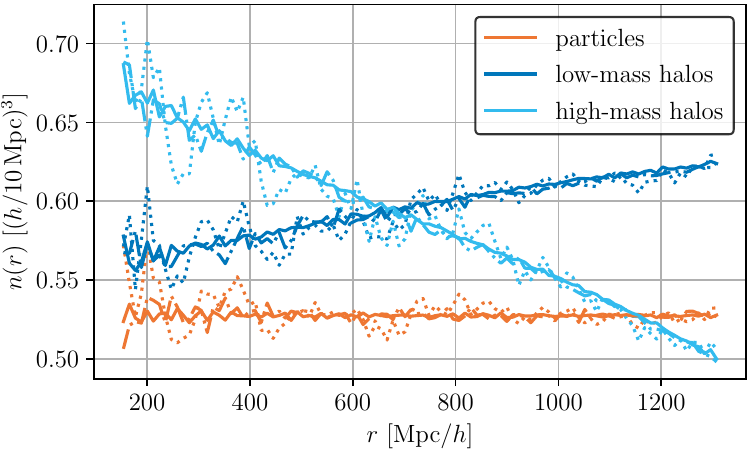}
    \caption{Number density of objects in the particle and halo data catalogues (dotted), and of corresponding Poisson-sampled random catalogues (dash-dotted) and glass catalogues (solid). The curves represent the number of objects per $(10\,\mathrm{Mpc}/h)^3$ within spherical shells of thickness $\Delta r \approx 12\,\mathrm{Mpc}/h$.}
    \label{fig:glass_random_n_of_r}
\end{figure}

In Fig.~\ref{fig:glass_random_n_of_r} the $n(r)$ measured from each data catalogue (dotted curves) are plotted together with the $n(r)$ of corresponding Poisson-sampled catalogues (dash-dotted curve) and glass catalogues (solid curve) with $\alpha = 1$. The density distributions of the Poisson-sampled random catalogues as well as those of the glass-like catalogues are in good agreement with the measured $n(r)$ of the data. The $n(r)$ of the Poisson-sampled catalogues fluctuate more strongly than the $n(r)$ of the glass-like random catalogues.

\begin{figure}

    \includegraphics[width=\columnwidth]{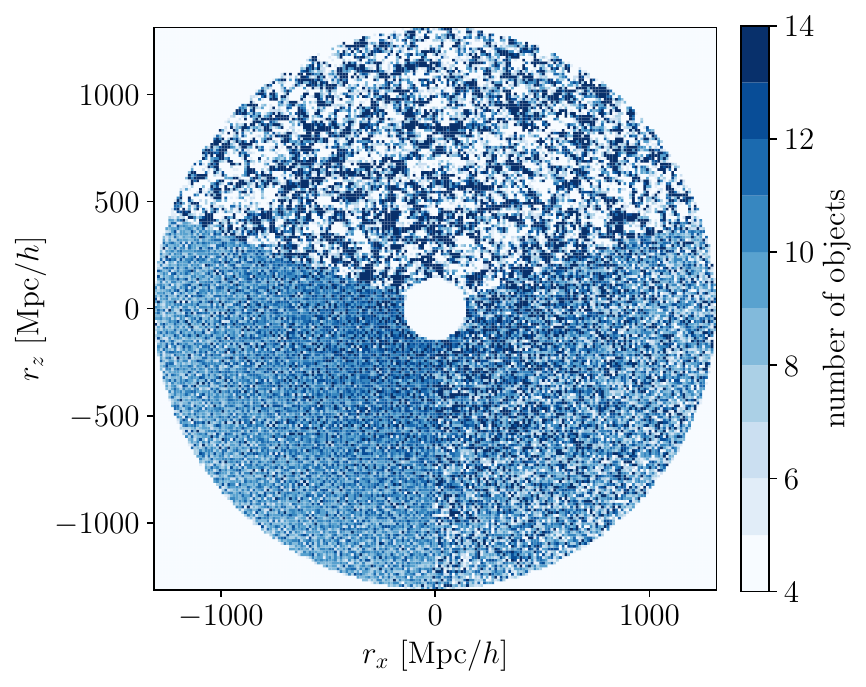}
    \caption{Projected number of objects in central slices with thickness $\Delta y = 100\,\mathrm{Mpc}/h$ through the high-mass halo light cone (top), a high-mass halo Poisson-sampled random catalogue with $\alpha = 1$ (bottom right) and a corresponding glass catalogue (bottom left).}
    \label{fig:high_slice}
\end{figure}

In Fig.~\ref{fig:high_slice} we show the projected number density of central slices with a thickness of $\Delta y = 100\,\mathrm{Mpc}/h$ through the high-mass halo light cone (top), and a corresponding Poisson-sampled random catalogue (bottom right) and glass-like random catalogue (bottom left) with $\alpha = 1$. The glass catalogue appears smoother and exhibits smaller fluctuations than the Poisson-sampled random catalogue on intermediate to large scales, while still reproducing the $n(r)$ of the data catalogue.

\begin{figure}

    \includegraphics[width=\columnwidth]{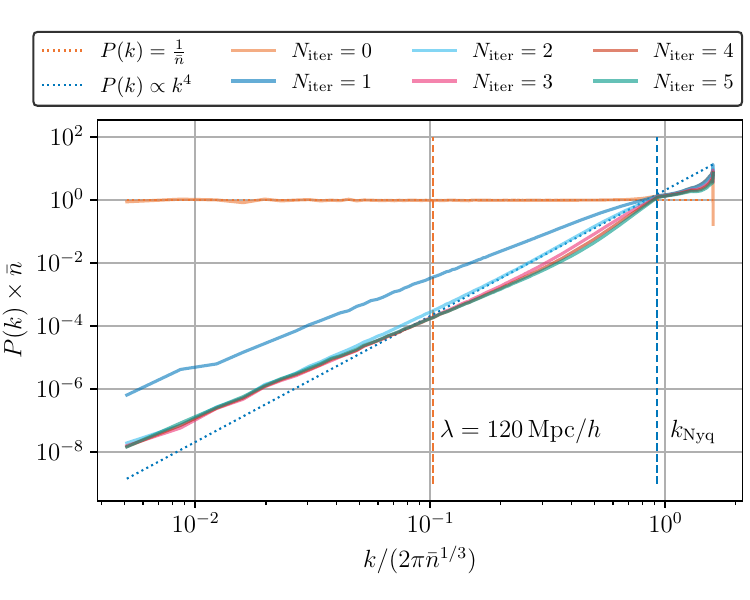}
    \caption{Power spectrum of the particle catalogue in the full box with $N_\mathrm{grid} = 512$, for different numbers of Zeldovich iterations $N_\mathrm{iter}$. Before applying any Zeldovich displacement the power spectrum is equal to the Poisson shot noise, $P(k) = 1/\bar{n}$. After one Zeldovich iteration the power on scales larger than the average inter-particle separation approaches $P(k) \propto k^4$. The blue dashed vertical line marks the Nyquist wave number $k_\mathrm{Nyq}$, the orange dashed vertical line marks the wavelength corresponding to the maximum scale of interest in our LS estimator of the 2PCF, $\lambda = d_\mathrm{max} = 120\,\mathrm{Mpc}/h$.}
    \label{fig:power}
\end{figure}

The number density of the particle catalogue is conserved, which means that the objects in the Poisson-sampled random catalogues and the glass-like catalogues are distributed uniformly in the periodic box (before removing the objects outside of the survey volume). We show the power spectrum of these catalogues estimated with {\sc Pylians} using a $512^3$ grid in Fig.~\ref{fig:power}. The case without any Zeldovich iterations, $N_\mathrm{iter} = 0$, corresponds to the Poisson-sampled catalogue, and the cases with $N_\mathrm{iter} > 0$ correspond to the glass-like catalogue after $N_\mathrm{iter}$ Zeldovich iterations. The Poisson-sample has a shot noise power spectrum, $P(k) = 1/\bar{n}$, as expected. The Zeldovich iterations decrease the power on scales larger than the average inter-particle separation.
The Nyquist wave number is marked with the blue dashed vertical line. The orange dashed vertical line marks the wave number that corresponds to fluctuations with wavelength equal to the maximum separation considered in the LS estimate of the 2PCF, $\lambda = d_\mathrm{max} = 120\,\mathrm{Mpc}/h$. Fluctuations with wavelengths much larger than that are expected to contribute only weakly to the correlation function and its variance below $d_\mathrm{max}$. We find that for $k$ above that wave number and below the Nyquist wave number the power spectrum of the glass-like catalogue is very close to $P(k) \propto k^4$ after $N_\mathrm{iter} = 2$ Zeldovich iterations, which is the result for a glass distribution.

\section{Results}
\label{sec:results}
\subsection{2PCF multipoles}

We estimate the full three-dimensional 2PCF of all possible combinations of data catalogues (DD), with 500 bins in $-1 \leq \mu \leq 1$ and 25 bins in $0\,\mathrm{Mpc}/h \leq d \leq 150\,\mathrm{Mpc}/h$ according to Eq.~\eqref{eq:ls}. This leads to $d$-bin widths of $\Delta d = 6\,\mathrm{Mpc}/h$. For each DD pair, we estimate the 2PCF using different types of random catalogues (C), which can be either Poisson-sampled random catalogues (R) or glass-like random catalogues (G), and different values of $\alpha$. For each of these cases, we perform 20 independent estimates of the 2PCF, using independent pairs of the same type of random catalogue. 
From each full 2PCF estimate, we  proceed to estimate the monopole, dipole and quadrupole ($\ell=0$, $\ell=1$, $\ell=2$) following Eq.~\eqref{eq:multipole}. 

\begin{figure}

    \includegraphics[width=\columnwidth]{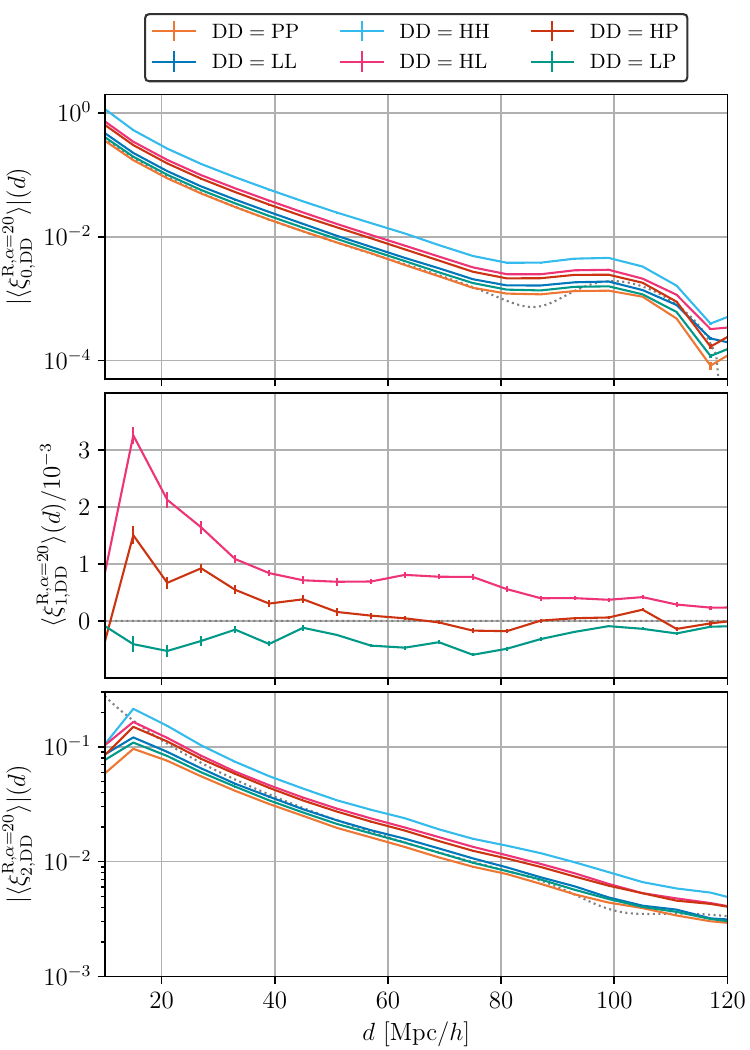}
    \caption{The sample mean monopole, dipole and quadropole of the LS estimates of the particle (P), low-mass halo (L) and high-mass halo (H) auto- and cross-correlations using 20 pairs of independent Poisson-sampled random catalogues with $\alpha = 20$. We show the absolute values of the even multipoles; the monopole becomes negative at $d \gtrapprox 117\,\mathrm{Mpc}/h$, while the quadruopole is always negative in the depicted range of $d$. For the dipole, only the cross-correlations are shown, as the autocorrelation dipoles vanish. The errorbars indicate the standard deviation in each bin. the The dotted grey lines are the linear theory predictions for the particle autocorrelation multipoles at the effective redshift $\bar{z} = 0.364$ produced with {\sc coffe}.}
    \label{fig:multipoles}
\end{figure}

Then, the sample mean $\langle\xi^{\mathrm{C},\alpha}_{\ell,\mathrm{DD}}(d)\rangle$ within each bin of $d$ is estimated from the 20 individual 2PCF multipole estimates according to

\begin{equation}
    \langle \xi^{\mathrm{C},\alpha}_{\ell,\mathrm{DD}}\rangle(d) \equiv \frac{1}{20}\sum_{i=0}^{20} \xi^{\mathrm{C},\alpha}_{i,\ell,\mathrm{DD}}(d)\,,
\end{equation}

where the $\xi^{\mathrm{C},\alpha}_{i,\ell,\mathrm{DD}}(d)$ are the individual LS estimates of the 20 2PCFs using the same type of random catalogue and the same value of $\alpha$. Then we estimate the sample standard deviation $\sigma^{\mathrm{C},\alpha}_{\ell,\mathrm{DD}}(d)$ as:

\begin{equation}
    \sigma^{\mathrm{C},\alpha}_{\ell,\mathrm{DD}}(d) \equiv \sqrt{\frac{1}{19}\sum_{i=0}^{20}(\xi^{\mathrm{C},\alpha}_{i, \ell,\mathrm{DD}}(d)-\langle \xi^{\mathrm{C},\alpha}_{\ell,\mathrm{DD}} \rangle(d))^2}\,,
\end{equation}
which is the square root of the sample variance estimator.
Fig.~\ref{fig:multipoles} shows the sample mean of the multipoles of the particle (P), low-mass halo (L) and high-mass halo (H) auto- and cross-correlation multipoles using the Poisson-sampled random catalogues with $\alpha=20$, which is understood to be an estimate close to the true 2PCFs of the respective data catalogues. For the even multipoles the absolute values are plotted. The monopole is negative for values of $d \gtrapprox 117 \mathrm{Mpc}/h$, and the quadrupole is always negative in the range of $d$ shown in the Figure. We show here only the dipoles of the cross-correlations, as the autocorrelations have a vanishing dipole.
The errorbars indicate the standard deviations in each bin of $d$. We additionally plot the theoretical prediction for the matter autocorrelation at the effective redshift of the particle catalogue, $\bar{z} = 0.364$ produced by {\sc coffe} \citep{Tansella:2018sld}, which includes all linear order relativistic effects in the full-sky two-point correlation function. 

The monopole signal of the particle autocorrelation agrees well with the theoretical linear prediction for the matter autocorrelation. The BAO peak in the correlation of the particle catalogue is broader than the one of the linear prediction, which is an expected result from nonlinear clustering in redshift space \citep{McCullagh:2012qy, McCullagh:2014jsa}.

The dipole signal is an expected result for redshift space cross-correlations of differently biased tracers \citep{Bonvin:2013ogt}, such as low- and high-mass halos (in this context, the linear bias $b$ is a quantity which encapsulates how much stronger a field $\delta_\mathrm{H}$ is clustered in comparison to the matter field $\delta_\mathrm{m}$ via the relation $\delta_\mathrm{H} = b\delta_\mathrm{m}$, from which it is evident that the linear bias is also expressed through the amplitude of the 2PCF monopole compared to the 2PCF monopole of matter). Here, linear theory predicts that the amplitude of the dipole signal is positively affected by large differences in the linear bias between the correlated populations. The linear biases of the high-mass and low-mass halo catalogues can be estimated from their monopoles and the particle autocorrelation monopole:

\begin{equation}
    b = \sqrt{\frac{\xi_{0,\mathrm{DD}}}{\xi_{0,\mathrm{PP}}}}\,,
\end{equation}
where DD is either HH for the high-mass halos or LL for the low-mass halos. Alternatively, one can divide the respective halo-particle cross-correlation monopole by the particle autocorrelation monopole to get a separate estimate of the linear bias. Using both of these methods, we derive a linear bias $b_\mathrm{HH}\approx 1.76$ for the high-mass halo catalogue and $b_\mathrm{LL}\approx 1.47$ for the low-mass halo catalogue. The linear bias of the particle catalogue is per definition $b_\mathrm{PP} = 1$, so the linear bias difference between the high-mass halo catalogue and the particle catalogue is larger than the linear bias difference between the high-mass halo catalogue and the low-mass halo catalogue. Taking only this linear bias difference into account, the expected dipole amplitude of the cross-correlation between the high-mass halos and the particles should be larger than that of the cross-correlation between high-mass halos and low-mass halos, but the opposite is true.

There is an additional contribution to the redshift-space 2PCF dipole from the so-called evolution bias \citep{Maartens:2021dqy}
\begin{equation}
\label{eq:fevo}
    f^\mathrm{evo} \equiv \frac{\mathrm{d}\mathrm{ln}n(z)}{\mathrm{d}\mathrm{ln}a}\,,
\end{equation}
where $n(z)$ is the comoving number density of the catalogue and $a$ is the cosmic scale factor, $a=1/(1+z)$. The evolution bias quantifies the evolution of a population's comoving number density with respect to redshift due to e.g. merging -- as can be seen in Fig.~\ref{fig:n_of_r} the number density of the high-mass halos decreases with redshift, while the number density of low-mass halos increases. Similar to the linear bias, a large evolution bias difference leads to a predicted increase of the dipole amplitude. We estimate the evolution bias of the halo catalogues according to Eq.~\eqref{eq:fevo}, and get $f^\mathrm{evo}_\mathrm{HH} \approx 0.85$ and $f^\mathrm{evo}_\mathrm{LL} \approx -0.28$. Again, per definition the evolution bias of the particle catalogue is zero, so indeed there is a larger bias difference between the high-mass halo catalogue and the low-mass halo catalogue, which might explain why the dipole amplitude is larger for the high-mass halo low-mass halo cross correlation. However, plugging these values into the theoretical model of {\sc coffe} (not shown here), we still find a disagreement with the dipole amplitude of our measurements.
We suspect that a contributing factor could be the unequal distribution of the two halo populations across different redshifts. While {\sc coffe} predicts the dipole signal at a fixed effective redshift, the dipole we observe is taken over a relatively large redshift range. At large redshifts, there are more low-mass halos and fewer high-mass halos, possibly contributing to the final dipole with a larger bias difference than what is inferred from the 2PCF monopoles taken across the whole redshift range. The exact theoretical modeling of this effect is beyond the scope of this paper and will be investigated in future work.

The linear biases are also expressed in the quadrupole estimates. While the linear theory prediction follows approximately the trend of the particle catalogue estimate, it does not match exactly. We suggest this can be attributed to non-linear contributions like the finger-of-God effect \citep{Jackson:1971sky}, which is an apparent elongation along the line of sight of the redshift-space tracer distribution close to the centers of clusters, induced by a Doppler shift from the velocity dispersion. The discrepancy between the linear theory prediction of \citet{Kaiser:1987qv} and measurements on scales below $\sim20\,\mathrm{Mpc}/h$ is well known and attempts at including this effect into the modeling have been made (see e.g. \citet{Peacock:2001gs, Scoccimarro:2004tg, Bianchi:2014kba, Bianchi:2016qen, BOSS:2016ntk}). In the quadrupole the finger-of-God effect acts opposite to the redshift space distortions on larger scales which appear as a flattening of the tracer distribution due to the Doppler shift induced by coherent infall velocities of the tracers. Indeed, in Fig.~\ref{fig:multipoles} the estimated quadrupole of the particle catalogue 2PCF is suppressed with respect to the theory prediction on scales below $\sim 80\,\mathrm{Mpc}/h$, with the difference becoming larger going to smaller scales, which supports the assumption of non-linear velocity contributions playing a role in explaining this discrepancy.

\subsection{Bias and standard deviation of the 2PCF multipole estimates}
We now quantify the accuracy and precision of the different LS estimates. In this context, we define the bias $\Delta\xi_{\ell,\mathrm{DD}}^{\mathrm{C},\alpha}(d)$ of each sample mean 2PCF multipole with respect to the corresponding estimate using the Poisson-sampled random catalogues with $\alpha=20$:

\begin{equation}
    \Delta\xi_{\ell,\mathrm{DD}}^{\mathrm{C},\alpha}(d) \equiv \langle \xi_{\ell,\mathrm{DD}}^{\mathrm{C},\alpha}\rangle(d) - \langle \xi_{\ell,\mathrm{DD}}^{\mathrm{R},\alpha=20}\rangle(d)\,,
\end{equation}
which contains information on the accuracy of the 2PCF estimate. The precision of each estimate is quantified by the ratio of its standard deviation to the standard deviation of the corresponding measurement using the Poisson-sampled random catalogue with $\alpha = 20$:

\begin{equation}
    \mathcal{R}^{\mathrm{C},\alpha}_{\ell,\mathrm{DD}}(d) \equiv \frac{\sigma^{\mathrm{C},\alpha}_{\ell,\mathrm{DD}}(d)}{\sigma^{\mathrm{R},\alpha=20}_{\ell,\mathrm{DD}}(d)}\,,
\end{equation}

\begin{figure}
\includegraphics[width=\columnwidth]{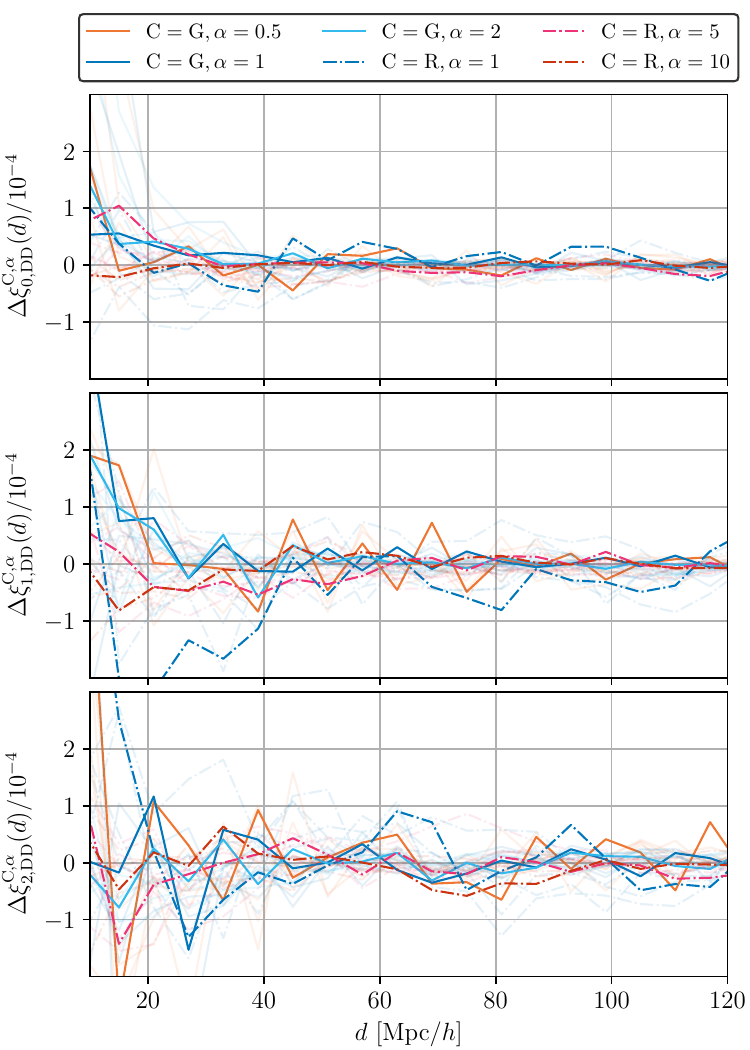}
\caption{Bias of the multipoles of the estimated cross-correlation between the high-mass halo catalogue and the particle catalogue with different choices of random catalogues and values of $\alpha$. Solid lines represent results using glass-like catalogues, while dash-dotted lines represent results from using Poisson-sampled random catalogues. The biases of the LS estimates of the other data catalogue pairs are plotted with a semi-transparent line style.}
\label{fig:biases}
\end{figure}

The biases and standard deviations of the 2PCF estimates of the different data catalogue pairs $\mathrm{DD}\in\{\mathrm{HH},\mathrm{HL},\mathrm{LL},\mathrm{HP},\mathrm{LP},\mathrm{PP}\}$ are similar when the particular choice of random catalogue and $\alpha$ is the same. 
In Fig.~\ref{fig:biases} we plot the bias of the multipoles of cross-correlation between the high-mass halo catalogue and the particle catalogue for different choices of random catalogues and values of $\alpha$.
The biases of the LS estimates of the other data catalogue pairs are plotted with a semi-transparent line style to demonstrate the expected range of the bias depending on the used data catalogues. The absolute value of the bias for all measurements stays well below $10^{-4}$ for all multipoles of interest and separations $d > 20\,\mathrm{Mpc}/h$. Increasing the value of $\alpha$ leads to less fluctuations in the bias for all multipoles irrespective of the chosen type of random catalogue, and to a smaller span between the minimum and maximum bias across the different data catalogue combinations. At small scales, $d<20\,\mathrm{Mpc}/h$, the bias
fluctuates more strongly, e.g. for the monopole using the glass-like catalogue with $\alpha=0.5$, it increases up to $2\times10^{-4}$. We suspect this is mainly due to a more noisy estimate of the 2PCF multipoles on smaller scales, as on these scales there are fewer pairs to be counted. 

It is expected that the measurements using the glass-like catalogues behave like the measurements using the Poisson-sampled random catalogues with the same $\alpha$ at scales below the average inter-particle separation. Nevertheless, most evident for the monopole estimates, the glass results start to disagree systematically with the results from using the Poisson-sampled random catalogues at small scales, $d<20\,\mathrm{Mpc}/h$, as the bias increases to values of $\sim 2\times 10^{-4}$ at $d=10\,\mathrm{Mpc}/h$ for any of the glass-like catalogues used in this work. We suspect that this additional bias at small scales can be attributed to the limitations of the Zeldovich approximation, which requires a discrete grid over which the density distribution is smoothed. If the size of the grid cells is not significantly smaller than the average inter-particle separation, a bias is introduced to the LS estimate of the 2PCF multipoles on scales below the grid cell size (see also Appendix~\ref{app:convergence}).

\begin{figure}
    \includegraphics[width=\columnwidth]{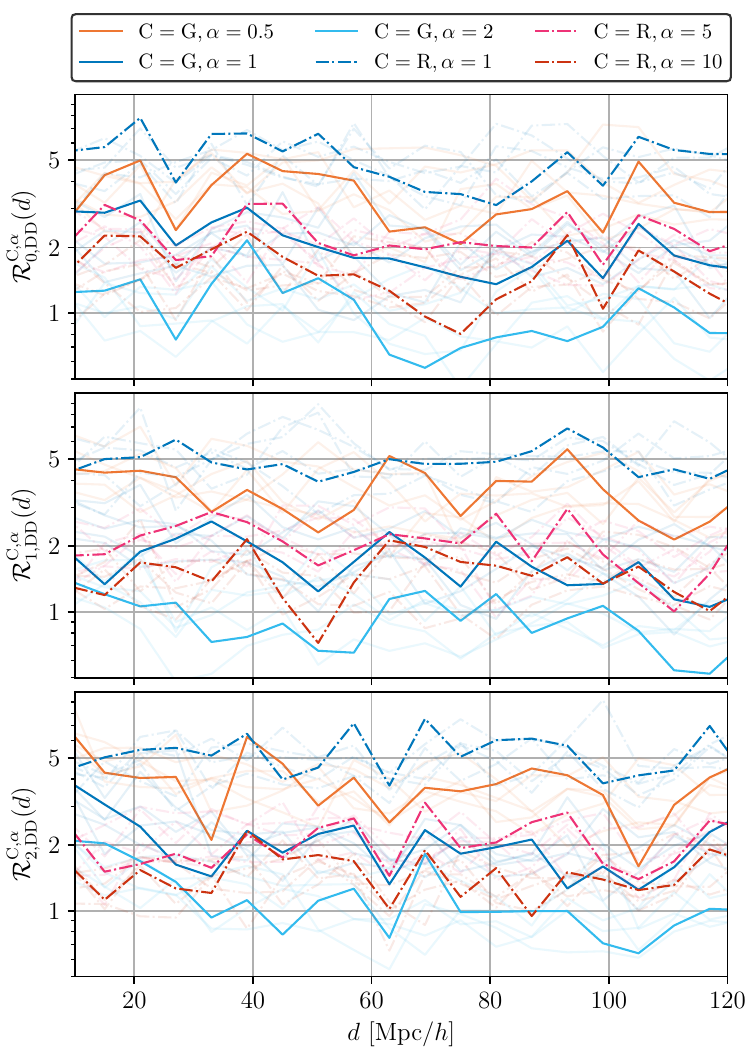}
    \caption{Standard-deviation ratio of estimated multipoles of the cross-correlation between the high-mass halo catalogue and the particle catalogue for different choices of random catalogues and values of $\alpha$ in Fig.~\ref{fig:stds}. Solid lines represent results from using glass-like random catalogues, while dash-dotted lines represent results from using Poisson-sampled random catalogues. The standard-deviation ratios of the LS estimates of the other data catalogue pairs are plotted with a semi-transparent line style.}
    \label{fig:stds}
\end{figure}

We plot the standard-deviation ratio of the estimated multipoles of the cross-correlation between the high-mass halo catalogue and the particle catalogue for different choices of random catalogues and values of $\alpha$ in Fig.~\ref{fig:stds}. The standard-deviation ratios of the LS estimates of the other data catalogue pairs are plotted with a semi-transparent line style to demonstrate the expected range of the standard-deviation ratio depending on the used data catalogues. The glass-like catalogues outperform the Poisson-sampled random catalogues significantly and to similar degree on all scales considered in this work. At $\alpha=1$ the standard-deviation ratio using the glass-like random catalogues is only $ \mathcal{R}_{\ell,\mathrm{HP}}^{G,\alpha=1}(d)\sim 2$, while for the Poisson-sampled random catalogues, it is $ \mathcal{R}_{\ell,\mathrm{HP}}^{G,\alpha=1}(d) \sim 5$, i.e. in this case the estimate using the glass catalogue is more than twice as precise as the estimate using the Poisson-sampled random catalogue. The standard deviation using the glass-like catalogue with $\alpha=2$ is almost the same as using the Poisson-sampled random catalogue with $\alpha=20$, which is a significant improvement. This holds approximately for all auto- and cross-correlations considered in this work, with minor fluctuations.

\subsection{Scaling of the standard deviation with $\alpha$}
\label{sec:scaling}

\begin{figure}
    \includegraphics[width=\columnwidth]{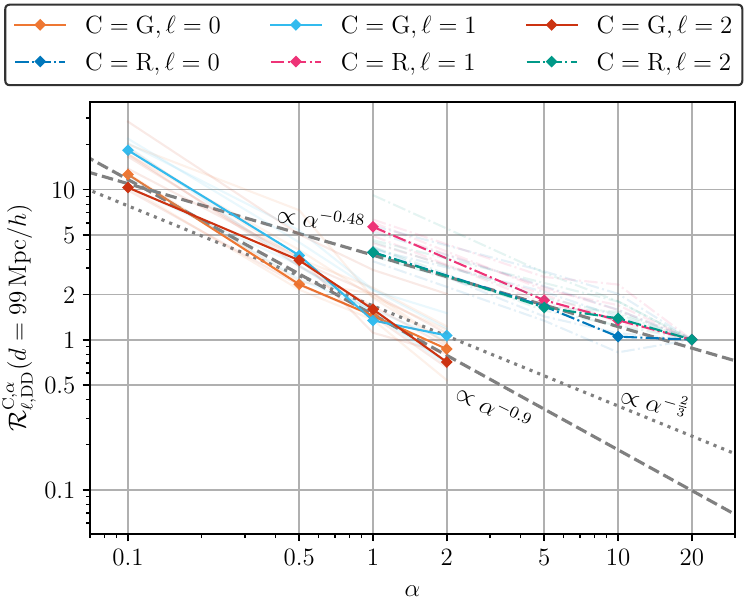}
    \caption{Standard-deviation ratio of the estimated multipoles of the cross-correlation between the high-mass halo catalogue and the particle catalogue in the bin centered at $d = 99\,\mathrm{Mpc}/h$ using glass-like random catalogues (solid curves) and Poisson-sampled random catalogues (dash-dotted curves), versus the value of $\alpha$. The transparent lines show the results for the LS estimates of the other data catalogue pairs. We find different power-law behaviours for the LS estimates of the 2PCF monopole, depending on the type of random catalogue used: $\mathcal{R}_{0,\mathrm{HP}}^{\mathrm{G},\alpha}(d = 99\,\mathrm{Mpc}/h) \propto \alpha^{-0.9}$ and $\mathcal{R}_{0,\mathrm{HP}}^{\mathrm{R},\alpha}(d = 99\,\mathrm{Mpc}/h) \propto \alpha^{-0.48}$ (dashed grey lines). The power law behaviour of the other two multipoles is similar. Additionally, the grey dotted line shows the expected power law for LS estimates utilising glass-like catalogues according to the considerations in Sec.~\ref{sec:variance}.}
    \label{fig:stds_vs_alpha}
\end{figure}

We now have a closer look at the scaling of the standard deviation of the LS estimate with increasing numbers of objects in the Poisson-sampled or glass-like catalogues, respectively.
To this end we plot the standard-deviation ratio of the estimated multipoles of the cross-correlation between the high-mass halo catalogue and the particle catalogue in the bin centered at $d = 99\,\mathrm{Mpc}/h$ against $\alpha$, depending on the chosen type of random catalogue, in Fig.~\ref{fig:stds_vs_alpha}. Again, the transparent lines show the results for the LS estimates of the remaining data catalogue pairs. Here, we also include results from using a glass-like catalogue with $\alpha = 0.1$, which have a significantly higher median standard deviation than the other catalogues with larger $\alpha$. From the considerations in Sec.~\ref{sec:variance} we expect different power-law behaviours of the standard deviation ratio with increasing $\alpha$, depending on the type of random catalogues used in the LS estimate. We fit power laws to the data (grey dashed lines) and find that the standard-deviation ratio of the monopole estimate using the glass-like catalogues behaves like $\mathcal{R}_{0,\mathrm{HP}}^{\mathrm{G},\alpha}(d = 99\,\mathrm{Mpc}/h) \propto \alpha^{-0.9}$ while for the Poisson-sampled random catalogues the power law is less steep: $\mathcal{R}_{0,\mathrm{HP}}^{\mathrm{R},\alpha} (d = 99\,\mathrm{Mpc}/h) \propto \alpha^{-0.48} \approx \alpha^{-0.5}$. Hence, the larger the choice of $\alpha$, the bigger the advantage of the glass-like catalogues. A similar behaviour is found for the other two multipoles investigated in this work. The grey dotted line shows the expected power law for LS estimates utilising glass-like catalogues according to the considerations in Sec.~\ref{sec:variance}. The measured power law for the LS estimate utilising glass-like catalogues disagrees slightly with the expected result, but is within the range of fluctuations from the measurements between all the data catalogue combinations.

\section{Conclusions}
\label{sec:conclusion}
Building on top of the ideas outlined in \cite{Davila-Kurban:2020eph}, we developed the publicly available code {\sc grlic}, which can be used to generate glass-like point distributions with radially-dependent number densities. The main application of this code is the generation of random catalogues for pair-count estimates of $N$-point correlation functions of e.g. galaxy-clustering data on cosmological light cones. 

The commonly used LS estimator relies on the assumption that a random catalogue with a very large number density is used. If this assumption holds the LS estimator is unbiased, and its variance is dominated by the variance of the random catalogue. Hence, it is common practice to generate Poisson-sampled random catalogues that mimic the radially-dependent number density and contain a factor of order $\sim10$ to $\sim100$ more objects than the data catalogues.

An alternative approach has been explored by \cite{Davila-Kurban:2020eph}. Glass-like catalogues can be generated from Poisson-sampled random catalogues using the Zeldovich approximation, and these catalogues exhibit considerably less power on scales above the average inter-particle distance, and hence require a lower number density than the usual Poisson-sampled random catalogues to achieve a similar variance in the LS estimator. This method has been applied to data catalogues with uniform background number densities, e.g. simulated galaxies within a periodic box, on an equal-time hypersurface. Glass-like catalogues are a convenient choice here, because they have already been investigated on their possible use for generating pre-initial conditions within cosmological N-body simulations, which evolve particles under the influence of gravity on an equal-time hypersurface within a periodic box. For the first time, we show that it is possible to create glass-like catalogues that follow a given background density distribution on large scales but qualitatively preserve the glass-like properties on small scales. This is a big advantage to periodic crystals which will have discontinuities if their large scale density is to be inhomogeneous.

We apply this idea to radially-dependent distributions as they are common in galaxy or halo catalogues with selection functions and verify that the non-uniform glass-like random approach gives similarly promising results as they have been reported in \cite{Davila-Kurban:2020eph}. For this purpose, we estimate the auto- and cross-correlations of three sets of simulated data catalogues using different Poisson-sampled and glass-like random catalogues with various numbers of objects: one particle catalogue with constant comoving number density, one high-mass halo catalogue with decreasing comoving number density as the comoving distance increases, and one low-mass halo catalogue with an increasing comoving number density as the comoving distance increases. We extract the first three multipoles of the LS estimator of these 2PCFs and find that the glass-like catalogues generated in this way outperform the Poisson-sampled catalogues significantly. 

We find that no significant bias is introduced on most scales when using glass-like random catalogues. Only on scales below $d = 20\,\mathrm{Mpc}/h$ there is a slight increase of the bias with our fiducial setup, but relative to the signal, it is still very small. The tests performed in Appendix~\ref{app:convergence} hint to the idea that this additional small bias is caused by inaccuracies in the Zeldovich approximation due to a limited grid resolution. Hence, it will be interesting to explore the results using an N-body $\mathrm{PM}^3$ scheme, which works just like the Zeldovich approximation on large scales, but uses direct particle-particle force calculations on smaller scales. It is expected that this approach will be less biased, but the required additional computing resources might not make up for the reduction of the bias.

Currently, the variance in the estimate of the cosmic 2PCF of galaxies is dominated by the sample variance. Given the ongoing technological advances for cosmological surveys we speculate that future surveys that cover a larger volume of the observable Universe and contain a higher density of tracers of the underlying density distribution will have reduced sample variance and hence might eventually warrant a need for a more precise estimator of the 2PCF, which can be acquired by gaining control of the variance arising from the statistical fluctuations in the random catalogues. This variance can be reduced by increasing the number of objects in the random catalogue.

Alternatively, glass-like catalogues can be used instead of Poisson-sampled random catalogues, as the LS estimates using the glass-like random catalogues have significantly reduced variance. We find a power-law behaviour of the standard deviation ratio at $d = 99\,\mathrm{Mpc}/h$, where the glass-like estimates of the monopole of the cross-correlation between the high-mass halos and the particles follow a steeper power law than the Poisson-sampled random estimates: $\mathcal{R}_{0,\mathrm{HP}}^{\mathrm{G},\alpha} (d = 99\,\mathrm{Mpc}/h) \propto \alpha^{-0.9}$ and $\mathcal{R}_{0,\mathrm{HP}}^{\mathrm{R},\alpha} (d = 99\,\mathrm{Mpc}/h) \propto \alpha^{-0.48}$. While we recover the expected behaviour for the standard deviation of the LS estimate using the Poisson-sampled random catalogues, the power-law slope for the glass-like random catalogues is steeper than the expected value of $-2/3$. We propose one possible explanation for this discrepancy: for the measurements with $\alpha < 1$ we suspect that the variance of the $R_iR_j$ term could become important and contribute to the power law with a different slope. For the glass distributions, the standard deviation of the $R_iR_j$ term is expected to decrease like $N_{R_i}^{-2/3}N_{R_j}^{-2/3} \approx N_{R_i}^{-4/3}$. The combined contributions of the $D_iR_j$ and $R_iR_j$ terms could then lead to a broken power law which only goes like $\alpha^{-2/3}$ for larger values of $\alpha$, and like $\alpha^{-4/3}$ for small values of $\alpha$. The value we measure is in between the expected values for the two dominant contributions, which could hint that indeed both contributions are relevant in the range of sampled values of $\alpha$. 

In general, slight deviations from the Poisson variance are expected even for the LS estimate utilising the Poisson-sampled catalogues. This is because the bin width and survey geometry affect the variance via the so-called edge terms (see e.g. \citet{Landy:1993yu, Keihanen:2019vst}). Since the width of the $d$-bins used in the 2PCF estimate affects its variance, it matters also when assessing the benefit of using glass-like random catalogues over Poisson-sampled random catalogues (see also Appendix~\ref{app:convergence}). If the bin size is chosen to be too small, i.e. much smaller than the average inter-particle separation, the variance of the estimate using glass-like catalogues is similar to the variance of the estimate using Poisson-sampled random catalogues, because in the glass-like catalogue the fluctuations are only suppressed on scales larger than the average inter-particle separation and fluctuations in the number counts between two small neighbouring bins are uncorrelated, just as they are in the Poisson-sampled random catalogue. As the bin width is increased, the correlated nature of the larger scale fluctuations in the glass-like catalogues makes the variance decrease faster than in the Poisson sample, whose fluctuations remain uncorrelated on all scales. This could provide another explanation for the unexpected power law-slope for the glass-like random catalogues: as we decrease $\alpha$, but keep the bin width fixed, the variance contribution from the bin width can become non-negligible such that the variance using glass-like random catalogues becomes more similar to the variance one gets when using Poisson-sampled random catalogues.

In future work, it would be interesting to perform a robust theoretical modeling of the variance of the LS estimator using glass-like catalogues, in a similar fashion as in \cite{Keihanen:2019vst}, to gain a better understanding of this scaling.

For the data catalogues used in this work, we find that the LS estimate using a glass-like catalogue with $\alpha=2$ is as precise the LS estimate using a Poisson-sampled random catalogue with $\alpha=20$. While this translates to a hundredfold speedup of the $\mathcal{O}(N^2)$ computation of the $R_1R_2$ pair-counts, increasing $\alpha$ further would make the advantage of the glass-like catalogues even more pronounced, due to the steeper power law with the glass-like catalogues (even though we expect the power law slope to become closer to $-2/3$ for larger $\alpha$). 

Our results apply specifically to catalogues with survey volumes, number densities and estimator bin widths as specified in this work, but should translate to catalogues with different volumes and number densities if the $d$-bin width is adapted accordingly and if a well-resolved Zeldovich grid is computationally affordable.

As an example, consider a typical survey with lower number density, such as the SDSS DR14Q quasar catalogue \citep{SDSS:2017oqu} which contains around $80$ quasars per square degree at redshifts $0.9 < z < 2.2$ within a survey area of $2044\,\mathrm{deg}^{2}$. Given the redshift depth and the survey area, the volume $V$ that contains $N = 80\,\mathrm{deg}^{-2} \times 2044\,\mathrm{deg}^2 = 163520$ quasars is calculated, $V = 4\pi/3 (r(z=2.2)^3 - r(z=0.9)^3) f$, where $f$ is the fraction of the survey area with respect to the full sky, $f = 2044/ 41253$, and the comoving distances to each redshift are $r(z=2.2) \approx 3756.5\,\mathrm {Mpc}/h $ and $r(z=0.9) \approx 2115.4\,\mathrm {Mpc}/h $. From these one can estimate the observed number density of quasars in that redshift bin to be $N/V = n \approx 1.8\times10^{-5} h^3/\mathrm{Mpc}^3$, which is approximately 28 times lower than the number density of the particle catalogue investigated in this work. The average inter-particle separation is then $d_\mathrm{inter} \approx 38\,\mathrm{Mpc}/h$ and an improvement of the variance is expected if the $d$-bin width is chosen to be not much smaller than $d_\mathrm{inter}/\alpha$. The required box size for such a survey is given by twice the distance to the maximum redshift $z=2.2$ with an additional buffer zone of $\approx 400\,\mathrm{Mpc}/h$ on each side, which results in approximately $8400\,\mathrm{Mpc}/h$. Then, the required number of grid cells such that the cell size is approximately equal to one quarter of the average inter-particle separation is $N_\mathrm{grid} = 4\times8400/38 \approx 880$. The required resolution is similar to the resolution used in the paper, so it is reasonable to assume that creating glass-like catalogues for this case is efficient. Due to the smaller number densities in such a survey, the shot noise of a Poisson-sampled random catalogue at a fixed $\alpha$ would be larger. One could minimize the dominant Poisson variance contribution of the data-random pairs by choosing a larger $\alpha$ such that the number of random-random pairs with separations below the maximum scale of interest remains small enough to be computable in a reasonable amount of time. Creating a glass-like catalogue could reduce the variance contribution of the data-random pairs even further, and the results presented in this work suggest that this relative reduction is stronger for larger values of $\alpha$. The larger volume of the quasar survey makes the contribution from the sample variance smaller and thus increases the relative importance of the variance of the estimator itself. Therefore, creating glass-like catalogues for the estimation of the 2PCF in such a survey is expected to be worthwhile.

The $n(r)$-distributions of the catalogues investigated here vary by relatively small amounts over the survey depth. Introducing selection functions to the survey can lead to less trivial variations of the $n(r)$-distribution for each catalogue. Tests on artificial catalogues with stronger variations in $n(r)$ have shown that for these catalogues, glass-like catalogues produced with {\sc{grlic}} lead to similar improvements in the variance of the estimated 2PCF as was found for the catalogues investigated in this work.

We attempt to model the effects of a survey mask in Appendix~\ref{app:mask} and find that the estimator variance is reduced in a similar fashion when using glass-like random catalogues compared to using Poisson-sampled random catalogues for the specified mask, suggesting that the geometric contributions to the variance are sub-dominant. It is unclear whether a more complicated mask could eventually lead to the geometric contributions to the variance to dominate. When this happens, we expect that the advantage of using glass-like random catalogues over Poisson-sampled random catalogues might differ from what is reported in this work -- we suspect that the glass-like catalogues will exhibit a lower variance contribution from geometric effects if compared to Poisson-sampled catalogues, because in general the edges of the survey will be sampled more smoothly by the glass-like catalogue, since its fluctuations are suppressed with respect to the Poisson-sampled case on most scales. A detailed study of the effects of masks on our results is beyond the scope of this paper and is postponed to future work.

In practice, if {\sc{grlic}} is to be used for masked survey data, it is required that the underlying $n(r)$ distribution is derived from the survey beforehand by using Eq.~\eqref{eq:N(r)} and a version of Eq.~\eqref{eq:n(r)} that models the actual window function of the survey instead of setting it to 1 everywhere. From this underlying $n(r)$ distribution, an unmasked glass-like catalogue can be sampled with {\sc{grlic}}, and the survey mask is applied to the glass-like catalogue, e.g. by weighing each object accordingly, when computing the 2PCF estimate.

We suggest that the glass-like random catalogues will also prove to be useful for pair-count based estimates of higher order statistics such as the three-point correlation function, where the computation time scales like $\mathcal{O}(N^3)$. Creating glass-like random catalogues can also be beneficial for Fourier space analyses using e.g. the FKP estimator described by Feldman, Kaiser and Peacock \citep{Feldman:1993ky}, where high values of $\alpha$ are required to suppress the estimator variance. Here, the requirement of large $\alpha$ does not pose such a big problem for the computation time, because FFT algorithms can solve the discretised equations quickly on a grid regardless of the size of the random catalogue. However, being able to reduce the required $\alpha$ by using glass-like catalogues could be advantageous in some cases, e.g. if a speed up of the mass-assignment procedure used to assign the densities to the grid on which the FFT is perfomed is desired, or if there are memory limitations for storing the random catalogues.

The main limitations in creating the glass-like catalogues with {\sc grlic} are memory requirements for the discretised grid used in the Zeldovich approach. The grid needs to be finely spaced in order to avoid biasing of the 2PCF estimate at small scales. For surveys with maximum redshift $z_\mathrm{max} \approx 0.5$ and comoving number densities of $n \approx \,(h/ 10\,\mathrm{Mpc})^3$, a grid with $N_\mathrm{grid} = 1024$ is sufficient for unbiased measurements for $d \geq 20\,\mathrm{Mpc}/h$, but the demand for a larger number of grid cells increases as the maximum redshift or the comoving number density of the survey becomes larger in order to ensure a well-resolved Zeldovich grid with as few objects per grid cell as possible. In the current implementation of {\sc grlic}, the observer is put into the center of the discretized box, and the box size is adjusted to fit the maximum comoving distance of the objects within the data catalogue, effectively increasing the physical size of the grid cells. This will make creating glass-like catalogues for very deep pencil beam surveys with high number densities difficult. In the future, this aspect can be made more memory efficient: instead of creating a full spherical glass with the given $n(r)$ in a periodic cube, from which the final glass-like random catalogue is cut out, the survey volume can be placed into a minimum bounding box. Here, avoiding unwanted edge effects is more involved, because of the discontinuous number densities on the periodic boundaries of the bounding box. A transitional region between the survey volume and the bounding volume can be implemented to ensure stability within the survey volume. In the current implementation of {\sc grlic}, the bulk of the memory requirement is covered by the entities stored on the discretised grid. The background number density, the density contrast and the displacement field are stored using single precision floating point numbers. On a grid with $N_\mathrm{grid} = 1024$ these already require $\approx 4 + 4 + 13 = 21\,\mathrm{GB}$ of random-access memory. Again this could be circumvented by using $\mathrm{PM}^3$ N-body simulations which do not require high grid resolutions to accurately simulate the repellent forces on small scales. Nevertheless it is expected that in the near future, memory resources will become more available which will make the use of glass-like catalogues even more advantageous compared to using Poisson-sampled random catalogues. 




\section*{Acknowledgements}
The author thanks Julian Adamek for useful discussions and valuable feedback on the draft. This work is supported by the Swiss National Science Foundation. We used high-performance computing resources provided by the Swiss National Super\-computing Centre (CSCS) under pay-per-use agreements (project ID ``uzh34'').

\section*{Data Availability}
{\sc grlic}, which was used to create the glass-like catalogues on the light cone, is publicly available at \url{https://gitlab.uzh.ch/sebastian.schulz/grlic}.
The data catalogues and corresponding random catalogues, as well as the 2PCF measurements and their mean multipoles are accessible at \url{https://zenodo.org/record/7799509#.ZCxeW9JBwW0} (DOI: 10.5281/zenodo.7799509).

\section*{}
\textit{This is the preprint version of the manuscript which has been accepted for publication in MNRAS, with minor corrections. The official published version is found at \url{https://doi.org/10.1093/mnras/stad2868}.}



\bibliographystyle{mnras}
\bibliography{citations} 




\appendix

\section{Convergence Tests}
\label{app:convergence}
\begin{figure}

    \includegraphics[width=\columnwidth]{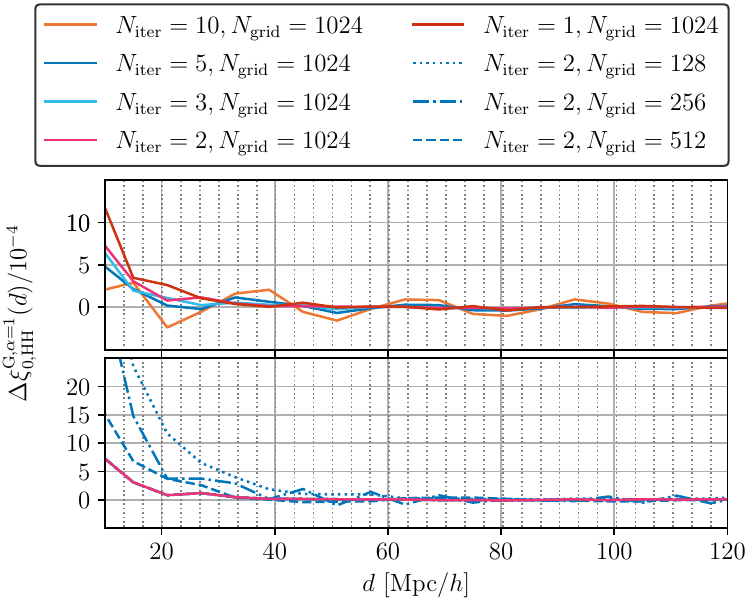}
    \caption{Bias of the monopole of the high-mass halo autocorrelation using glass-like random catalogues with $\alpha=1$ and $\Delta d = 6\,\mathrm{Mpc}/h$, created with different numbers of iterations of the Zeldovich approximation $N_\mathrm{iter}$ (top panel) and different numbers of grid cells $N_\mathrm{grid}$ (bottom panel). The grey dotted vertical lines represent the spacing between the grid cells in the fiducial grid with $N_\mathrm{grid} = 1024$. Increasing the number of iterations decreases the bias on scales below $20\,\mathrm{Mpc}/h$. After no more than ten iterations, the estimator bias oscillates around zero. Decreasing the number of grid cells increases the small scale bias of the estimates.}
    \label{fig:biases_convergence_monopole_25bins}
\end{figure}
\begin{figure}

    \includegraphics[width=\columnwidth]{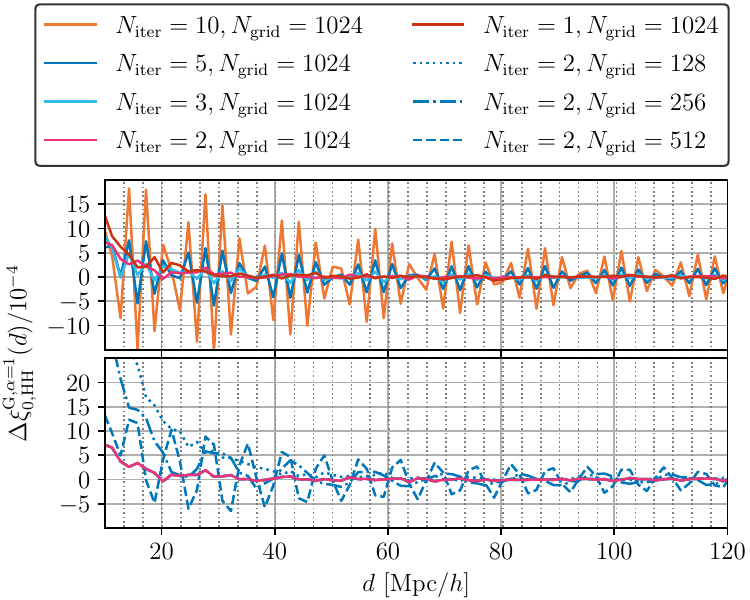}
    \caption{Bias of the monopole of the high-mass halo autocorrelation using glass-like random catalogues with $\alpha=1$ and $\Delta d = 1.5\,\mathrm{Mpc}/h$, created with different numbers of iterations of the Zeldovich approximation $N_\mathrm{iter}$ (top panel) and different numbers of grid cells $N_\mathrm{grid}$ (bottom panel). The grey dotted vertical lines represent the spacing between the grid cells in the fiducial grid with $N_\mathrm{grid} = 1024$. As the number of iterations increases, the monopole bias oscillates more and more around zero, with a period corresponding to the grid cell separation. This is attributed to grid alignment of the objects in the glass-like catalogue after too many iterations. Decreasing the number of grid cells increases the small scale bias of the estimates. For a fixed number of iterations $N_\mathrm{iter} = 2$, decreasing the grid resolution makes the oscillatory feature apparent, with periods whose dominant contribution is the respective grid cell separation.}
    \label{fig:biases_convergence_monopole}
\end{figure}
\begin{figure}

    \includegraphics[width=\columnwidth]{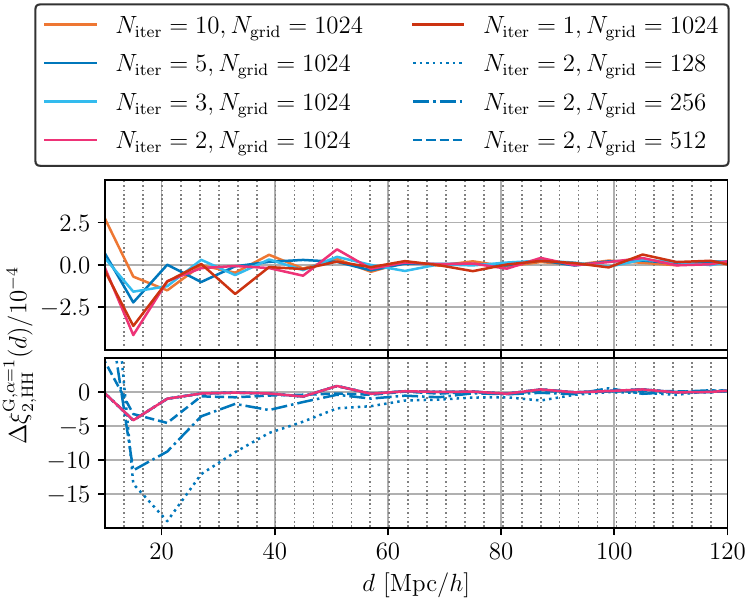}
    \caption{Bias of the quadrupole of the high-mass halo autocorrelation using glass-like random catalogues with $\alpha=1$ and $\Delta d = 6\,\mathrm{Mpc}/h$, created with different numbers of iterations of the Zeldovich approximation $N_\mathrm{iter}$ (top panel) and different numbers of grid cells $N_\mathrm{grid}$ (bottom panel). The grey dotted vertical lines represent the spacing between the grid cells in the fiducial grid with $N_\mathrm{grid} = 1024$. Decreasing the number of grid cells increases the small scale bias of the estimates.}
    \label{fig:biases_convergence_quadrupole}
\end{figure}

We test the convergence of our fiducial setup by varying the number of Zeldovich iterations $N_\mathrm{iter}$ as well as the number of grid cells $N_\mathrm{grid}$ used in the generation of glass-like catalogues with $\alpha=1$ for the LS estimate of the 2PCF multipoles for of the high-mass halo catalogue.

In general, each successive iteration of the Zeldovich displacement represents a higher order correction to the initially Poisson-sampled random catalogue, so already after one iteration, the Poisson noise is expected to be reduced by the largest amount due to the first-order correction. One can increase the number of iterations to add second- and higher-order corrections, but the question is whether at a certain point discretisation effects originating from the Zeldovich grid become evident, imposing a maximum on the acceptable number of iterations $N_\mathrm{iter}$.

Fig.~\ref{fig:biases_convergence_monopole_25bins} and Fig.~\ref{fig:biases_convergence_monopole} show the resulting bias for the 2PCF monopole using the fiducial number of 25 bins in $d$ corresponding to $\Delta d = 6\,\mathrm{Mpc}/h$ and using 100 bins in $d$ which translates to $\Delta d = 1.5\,\mathrm{Mpc}/h$, respectively. For the fiducial binning, we find that increasing the number of iterations (top panel) reduces the bias on scales below $20\,\mathrm{Mpc}/h$, but introduces an oscillation of the estimator bias around zero after no more than ten iterations. The oscillatory feature in the bias can be understood better when inspecting the results in Fig.~\ref{fig:biases_convergence_monopole}. Here, the oscillation becomes apparent already after as few as three iterations of the Zeldovich approximation. We suspect that after too many iterations, the objects in the catalogue eventually align with the grid. If that is the case, there are expected to be spikes in the LS estimate at separations that are equal to separations that are prominent in the the periodic grid point distribution. We add vertical dotted grey lines that are spaced with $\Delta = 3.346 \mathrm{Mpc}/h$ which corresponds exactly to the distance between each grid point with $N_\mathrm{grid}= 1024$. The period of the oscillations coincides very well with the spacing of the vertical grey lines. For $\Delta d = 1.5\,\mathrm{Mpc}/h$, we find that using more than three iterations is sub-optimal, as then the oscillations start to become noticeable.

Reducing the number of grid cells at $\Delta d = 6\,\mathrm{Mpc}/h$ (bottom panel of Fig.~\ref{fig:biases_convergence_monopole_25bins}) leads to an additional small scale bias in the monopole, and this additional bias becomes apparent at larger scales as the number of grid cells is reduced. Further, the bias starts to oscillate around zero as soon as $N_\mathrm{grid} = 256$. This is again more apparent for the binning with $\Delta d = 1.5\,\mathrm{Mpc}/h$ (see bottom panel of Fig.~\ref{fig:biases_convergence_monopole}). While with a grid resolution of $N_\mathrm{grid} = 1024$ for $N_\mathrm{iter} = 2$ there are no oscillations in the bias, decreasing the grid resolution to $N_\mathrm{grid} = 512$ leads to an apparent oscillation of the bias around zero with a period that corresponds to the new grid cell separation (i.e. twice the grid cell separation that corresponds to $N_\mathrm{grid} = 1024$). This hints to an interplay between the required grid resolution and the maximum possible number of iterations in the Zeldovich approximation in order to get an unbiased result, i.e. optimally, the grid should be as fine as possible to prevent biasing at small separations, and the number of Zeldovich iterations needs to be balanced such that it is high enough to give a converged glass with low bias at small separations, but small enough to prevent the oscillatory feature caused by discretisation of effects.

Fig.~\ref{fig:biases_convergence_quadrupole} shows the same test results as Fig.~\ref{fig:biases_convergence_monopole_25bins}, but for the quadrupole of the 2PCF.
The small scale bias is reduced in a similar fashion if the number of iterations is increased, but the oscillation around zero is not as present as for the monopole. Reducing the Zeldovich grid resolution introduces an additional bias on small scales, which grows larger and affects larger scales as the resolution decreases.

\begin{figure}

    \includegraphics[width=\columnwidth]{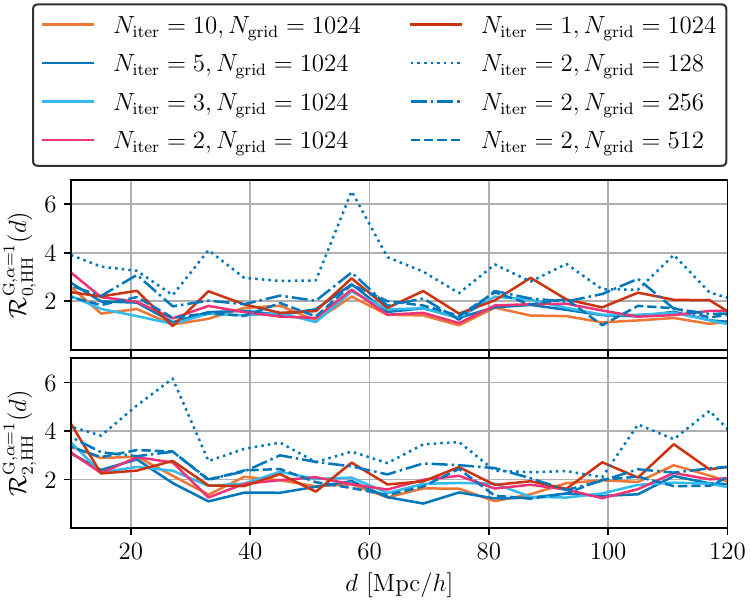}
    \caption{Standard-deviation ratio of the 2PCF monopole (top panel) and quadrupole (bottom panel) of the high-mass halo autocorrelation using glass-like random catalogues with $\alpha=1$, created with  different numbers of iterations of the Zeldovich approximation $N_\mathrm{iter}$ (solid curves) and different numbers of grid cells $N_\mathrm{grid}$ (dotted, dash-dotted and dashed curves). Increasing the number of iterations has negligible effect on the standard-deviation ratio. Decreasing the number of grid cells increases the standard-deviation ratio slightly.}
    \label{fig:stds_convergence}
\end{figure}

Fig.~\ref{fig:stds_convergence} shows the standard deviation ratio of the monopole and quadrupole of the high-mass halo autocorrelation estimate using the glass-catalogue with $\alpha=1$ with different $N_\mathrm{iter}$ and $N_\mathrm{grid}$. The standard deviation of the estimate is largely unaffected by different choices of $N_\mathrm{iter}$ and $N_\mathrm{grid}$. There is a slightly larger standard deviation if a smaller grid with $N_\mathrm{grid} = 128$ is used. This can be attributed to the fact that for such a very low number of grid cells, the creation of a glass-like catalogue is not working efficiently, as the grid cell size also imposes a lower limit down to which scales the power of the original Poisson-sampled catalogue can be suppressed. For better resolved grids we find that already after one iteration of the Zeldovich approximation, the standard deviation is basically converged.

Since the variance of the catalogue is well-suppressed already after $N_\mathrm{iter} = 1$, we suggest that a safe choice in general is not to use values of $N_\mathrm{iter}$ larger than $2$ in order to avoid discretisation effects while ensuring convergence of the glass. We note that \citet{Davila-Kurban:2020eph} used a larger number of iterations up to $N_\mathrm{iter} = 50$. They find that after as few as five iterations the variance is converged. A difference between their work and the work presented here is the definition of the density contrast that is the basis for calculating the Zeldovich displacement. \citet{Davila-Kurban:2020eph} choose to use a Gaussian kernel to smooth the particle distribution and estimate the densities on a grid. We use the cloud-in-cell mass-assignment scheme, and self-consistently interpolate the displacement field at the particle positions using a matching kernel. We suspect that using a Gaussian kernel to smooth the particle distribution from which the densities are estimated leads to an underestimation of the repulsive gravitational forces for each particle, because the amplitude of the displacement depends on the amplitude of the density contrast, which is smaller if the densities are smoothed beforehand. This could lead to a larger number of required Zeldovich iterations to reach a glass-like distribution.

\begin{figure}
    \includegraphics[width=\columnwidth]{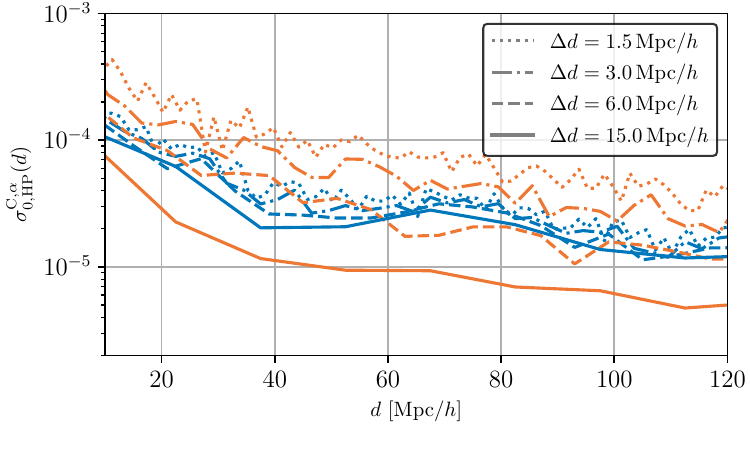}
    \caption{Standard deviation of the estimated 2PCF cross-correlation between the high-mass catalogue and the particle catalogue for different choices of $d$-bin widths $\Delta d$, using glass-like random catalogues with $\alpha = 2$ (orange curves) and Poisson-sampled random catalogues with $\alpha = 20$ (blue curves).}
    \label{fig:bin_convergence}
\end{figure}

We additionally plot the estimator variance for the cross-correlation between the high-mass halo catalogue and the particle catalogue for different choices of $d$-bin widths $\Delta d$ in Fig.~\ref{fig:bin_convergence}, using either glass-like random catalogues with $\alpha = 2$ (orange curves) or Poisson-sampled random catalogues with $\alpha = 20$ (blue curves). It is evident that the variance reduction when using glass-like random catalogues in the estimate is affected by the specific choice of $d$-bin width -- when using glass-like catalogues, the estimator variance decreases with increasing bin width, while for Poisson-sampled random catalogues, the estimator variance remains largely unaffected. Therefore, the advantage of using glass-like catalogues is larger the larger the width of the $d$-bins in relation to the average inter-particle separation. While the maximum bin width of $\Delta d = 15\,\mathrm{Mpc}/h$ tested here gives the best improvement of the variance, for our fiducial setup we use $\Delta d = 6\,\mathrm{Mpc}/h$, to have a finer sampling of the 2PCF estimates.

\section{Effects Of A Survey Mask}
\label{app:mask}

\begin{figure}
    \includegraphics[width=\columnwidth]{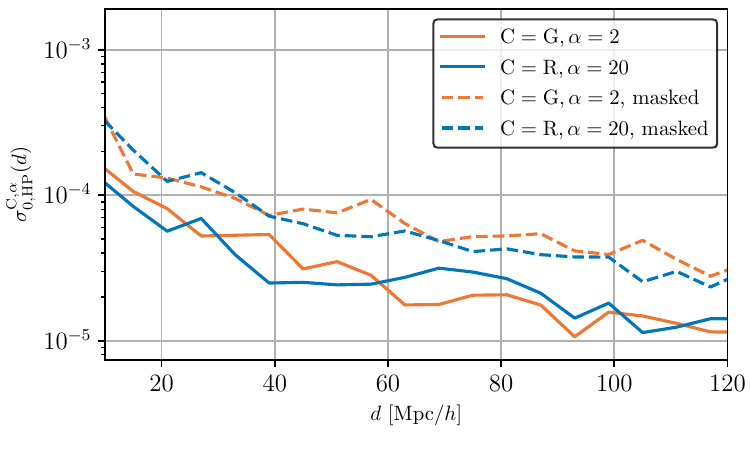}
    \caption{Standard deviation of the estimated 2PCF cross-correlation between the high-mass catalogue and the particle catalogue, for the fiducial case without any mask and the case after applying a mask, using a glass-like random catalogue with $\alpha = 2$ (orange curves) and a Poisson-sampled random catalogue with $\alpha = 20$ (blue curves).}
    \label{fig:masked_test}
\end{figure}

In order to test the effect of more complicated masks, we introduce a survey mask to the high-mass halo catalogue and the particle catalogue and estimate their cross-correlation and its variance using either masked glass-like random catalogues or masked Poisson-sampled random catalogues with different $\alpha$. The mask removes all objects with $\mu<-0.8$, $-0.6 < \mu<-0.4$, $-0.2 < \mu < 0$, $0.2 < \mu< 0.4$, $0.6 < \mu < 0.8$, as well as all remaining objects with $\varphi < -\frac{2}{3}\pi$, $-\frac{1}{3}\pi < \varphi < 0$,  $\frac{1}{3}\pi < \varphi < \frac{2}{3}\pi$. In Fig.~\ref{fig:masked_test} it is shown that the overall variance of the estimate increases when introducing the mask, but the glass-like catalogues provide the same advantage as in the unmasked case: using a $d$-bin width of $\Delta d = 6\,\mathrm{Mpc}/h$, the glass-like random catalogue with $\alpha=2$ has the same variance as the Poisson-sampled random catalogue with $\alpha = 20$. The increased variance of the masked estimate is attributed mainly due to a reduced number of objects available to the pair-counting algorithm.



\bsp	
\label{lastpage}
\end{document}